\documentclass[12pt,a4paper]{report}

\usepackage{thesis}
\usepackage{lmodern}
\usepackage[numbers,sort&compress]{natbib}
\usepackage[hidelinks]{hyperref}
\usepackage[left=2.5cm,top=2.5cm,right=2.5cm,bottom=2.5cm]{geometry}
\usepackage{nomencl}
\makenomenclature

\usepackage{tocloft}

\newcommand{\thesistitle}{Parameterized Algorithms for\\ Scalable Interprocedural Data-flow Analysis}
\newcommand{\thesisauthor}{Ahmed Khaled Abdelfattah Zaher}
\newcommand{\programname}{Computer Science and Engineering}
\newcommand{\departmentname}{Department of Computer Science and Engineering}
\newcommand{\thesisdate}{June 2023}
\newcommand{\signdate}{June 2023}
\newcommand{\thesisdegree}{MPhil }

\newcommand{\supervisorinfo}{Prof. Amir Kafshdar Goharshady, Thesis Supervisor\\Deparment of Computer Science and Engineering\\Department of Mathematics}
\newcommand{\depheadinfo}{Prof. Xiaofang Zhou}

\usepackage{xcolor}
\usepackage{fixltx2e}
\usepackage{tikz}
\usepackage{listings}
\usepackage{graphicx}
\usepackage{stmaryrd}
\usepackage{float}
\usepackage{enumerate}
\usepackage[mathcal]{euscript}
\usepackage{algpseudocode}
\usepackage[linesnumbered,ruled,vlined]{algorithm2e}
\usepackage{amsthm}
\usepackage{amsmath,amsfonts}
\usepackage{lmodern}
\usepackage[numbers,sort&compress]{natbib}
\usepackage[hidelinks]{hyperref}
\usepackage{microtype}
\usepackage{tabularray}
\usepackage{multirow}
\usepackage{hhline}
\usepackage{booktabs}

\newcommand{\summ}{\chi}
\newcommand{\qry}{\texttt{Q}}
\newcommand{\scqry}{\texttt{SCQ}}
\newcommand{\valid}{\texttt{IVP}}
\newcommand{\mivp}{\texttt{MIVP}}
\newcommand{\scv}{\texttt{SCVP}}
\newcommand{\mscvp}{\texttt{MSCVP}}
\newcommand{\meet}{\sqcap}
\newcommand{\zero}{\textbf{0}}
\renewcommand{\phi}{\varphi}
\newcommand{\bags}{\mathfrak{B}}
\newcommand{\subtree}{\overline{T}^{\downarrow}}
\newcommand{\descendants}{\overline{F}^\downarrow}
\usepackage{paralist}

\definecolor{codeblue}{rgb}{0.2,0.35,0.95}
\definecolor{codeblack}{rgb}{0,0,0}
\definecolor{codepurple}{rgb}{0.58,0,0.82}
\definecolor{backcolour}{rgb}{0.95,0.95,0.92}

\usepackage{csquotes}
\SetKwComment{Comment}{/* }{ */}

\newcommand{\TT}[1]{\texttt{#1}}
\renewcommand{\contentsname}{\centering Table of Contents}
\usepackage{sectsty}
\chapterfont{\centering}

\begin{document}
%\huge{Apply the changes of IVP and SCVP}
\pagenumbering{roman}
\pagestyle{plain}
\setcounter{page}{1}
\addcontentsline{toc}{chapter}{Title Page}
%!TEX program = xelatex
%!TEX root = ./thesis.tex
\thispagestyle{empty}
\null\vskip0.5in
\begin{center}
  \begin{LARGE}
    \thesistitle
  \end{LARGE}
  \vfill
  \vspace{20mm}

  by

  \vspace{4mm}

  \thesisauthor \\
  \vfill
  \vspace{20mm}

  A Thesis Submitted to\\
  The Hong Kong University of Science and Technology \\
  in Partial Fulfillment of the Requirements for\\
  the Degree of Master of Philosophy \\
  in \programname \\
  \vfill \vfill
  \thesisdate, Hong Kong
  \vfill
\end{center}

\vfill

\newpage
\addcontentsline{toc}{chapter}{Authorization Page}
%!TEX program = xelatex
%!TEX root = ./thesis.tex
\null\skip0.2in
\begin{center}
{\bf \Large \underline{Authorization}}
\end{center}
\vspace{12mm}

I hereby declare that I am the sole author of this thesis.

\vspace{10mm}

I authorize the Hong Kong University of Science and Technology to lend this
thesis to other institutions or individuals for the purpose of scholarly research.

\vspace{10mm}

I further authorize the Hong Kong University of Science and Technology to
reproduce the thesis by photocopying or by other means, in total or in part, at the
request of other institutions or individuals for the purpose of scholarly research.

\vspace{30mm}

\begin{center}
\underline{~~~~~~~~~~~~~~~~~~~~~~~~~~~~~~~~~~~~~~~~~~~~~~~~~~~~~~~~~~~~~~~~~~~~~~}\\
~~~~\thesisauthor \\
~~~~\signdate

\end{center}

\newpage
\addcontentsline{toc}{chapter}{Signature Page}
%!TEX program = xelatex
%!TEX root = ./thesis.tex
\begin{center}
{\Large \thesistitle}\\
\vspace{5mm}
by\\
\vspace{3mm}
\thesisauthor\\
\vspace{5mm}
This is to certify that I have examined the above \thesisdegree thesis\\
and have found that it is complete and satisfactory in all respects,\\
and that any and all revisions required by\\
the thesis examination committee have been made.
\end{center}

\vspace{15mm}

\begin{center}
\underline{~~~~~~~~~~~~~~~~~~~~~~~~~~~~~~~~~~~~~~~~~~~~~~~~~~~~~~~~~~~~~~~~~~~~~~~~~~~ }\\
\supervisorinfo\\
\end{center}

\vspace{15mm}
\begin{center}
\underline{~~~~~~~~~~~~~~~~~~~~~~~~~~~~~~~~~~~~~~~~~~~~~~~~~~~~~~~~~~~~~~~~~~~~~~~~~~~ }\\
\depheadinfo\\
Head, Department of Computer Science and Engineering
\end{center}

\vspace{5mm}
\begin{center}
\vspace{5mm}
\signdate
\end{center}

\newpage
\addcontentsline{toc}{chapter}{Acknowledgments}
%!TEX program = xelatex
%!TEX root = ./thesis.tex
\begin{center}
{\bf \Large \underline{Acknowledgments}}
\end{center}
\vspace{12mm}
I am immensely grateful to my advisor Amir Goharshady for giving me such excellent support and guidance, and for always thinking about what is best for me. He consistently presented me with research ideas and precious insights that developed my academic mindset and helped me become a better researcher, and he truly goes the extra mile to optimize my chances of a better career. I am further grateful for my personal relationship with him as a light-hearted and kind friend.

I am thankful to all the colleagues of our ALPACAS research group, who are not only talented researchers shaping the excellent research environment of our group but also great friends. I particularly thank Zhuo Cai, Giovanna Conrado, Soroush Farokhnia, Singh Hitarth, Pavel Hudec, Kerim Kochekov, Harshit Motwani, Sergei Novozhilov, Tamzid Rubab, Yun Chen Tsai, and Zhiang Wu.

I express gratitude to my parents Khaled Zaher and Gihan El Sawaf for being supportive of me and my pursuits, and to my lifelong friends Mohamed El-Damaty, Eyad Abu-Zaid, and Mustafa Elkasrawy.

Further gratitude goes to the good friends I made in Hong Kong, who helped me adapt to the city and feel at home. This includes Amr Arafa, Yuri Kuzmin, Ian Varela, Yipeng Wang, and Kenny Ma.

Finally, I am very grateful to Professors Andrew Horner and Jiasi Shen for kindly accepting to be on my thesis defence committee.

\newpage
\addcontentsline{toc}{chapter}{Table of Contents}
\tableofcontents

%\newpage
\addcontentsline{toc}{chapter}{List of Publications}
\chapter*{List of Publications}
\vspace{12mm}

$\bullet$ A. K. Goharshady and A. K. Zaher, ``Efficient interprocedural data-flow analysis using treedepth and treewidth,'' in Proceedings of the 24th International Conference on Verification, Model Checking, and Abstract Interpretation (VMCAI), 2023.\\~\\
$\bullet$ G.K. Conrado, A.K. Goharshady, K. Kochekov, Y.C. Tsai, A.K. Zaher, ``Exploiting the Sparseness of Control-flow and Call Graphs for Efficient and On-demand Algebraic Program Analysis,'' in Proceedings of the ACM SIGPLAN International Conference on Object-Oriented Programming Systems, Languages, and Applications (OOPSLA), 2023

\paragraph{Note:} Following the norms of theoretical computer science, co-authors are listed in alphabetical order. 

\newpage
\addcontentsline{toc}{chapter}{Abstract}
%!TEX program = xelatex
%!TEX root = ./thesis.tex
\begin{center}
{\Large \thesistitle}\\
\vspace{20mm}
by \thesisauthor\\
%\vspace{15mm}
\departmentname\\
%\vspace{10mm}
The Hong Kong University of Science and Technology
\end{center}
\vspace{8mm}
\begin{center}
Abstract
\end{center}
\par
\noindent
Data-flow analysis is a general technique used to compute information of interest at different points of a program and is considered to be a cornerstone of static analysis. In this thesis, we consider interprocedural data-flow analysis as formalized by the standard IFDS framework, which can express many widely-used static analyses such as reaching definitions, live variables, and null-pointer. We focus on the well-studied on-demand setting in which queries arrive one-by-one in a stream and each query should be answered as fast as possible. While the classical IFDS algorithm provides a polynomial-time solution to this problem, it is not scalable in practice. Specifically, it either requires a quadratic-time preprocessing phase or takes linear time per query, both of which are untenable for modern huge codebases with hundreds of thousands of lines. Previous works have already shown that parameterizing the problem by the treewidth of the program's control-flow graph is promising and can lead to significant gains in efficiency. Unfortunately, these results were only applicable to the limited special case of same-context queries.

In this work, we obtain significant speedups for the general case of on-demand IFDS with queries that are not necessarily same-context. This is achieved by exploiting a new graph sparsity parameter, namely the treedepth of the program's call graph. Our approach is the first to exploit the sparsity of control-flow graphs and call graphs at the same time and parameterize by both treewidth and treedepth. We obtain an algorithm with a linear preprocessing phase that can answer each query in constant time with respect to the input size. Finally, we show experimental results demonstrating that our approach significantly outperforms the classical IFDS and its on-demand variant.

\clubpenalty10000
\widowpenalty10000

\newpage
\pagenumbering{arabic}
\pagestyle{plain}
\setcounter{page}{1}
\renewcommand{\paragraph}[1]{\medskip\noindent\textbf{#1.}}
\chapter{Introduction} \label{sec:intro}

\paragraph{Static Program Analysis} Static program analysis is concerned with automatically finding bugs in programs. It is static in the sense that it achieves that goal by analyzing a program's source code without running it. Static analysis investigates questions about a program's behavior such as:
\begin{compactenum}[(i)]
\item Does the program use a variable \TT{x} before it is initialized?
\item Can the program have a null-pointer dereferencing?
\item For an expression \TT{e} that appears inside the body of a loop, does \TT{e}'s value depend on the loop iteration?
\end{compactenum}
The use of static program analysis in compiler optimization dates back more than half a century ago~\cite{DBLP:conf/comop/Allen70}. However, numerous other benefits have also emerged since. This includes aiding the programmer to find bugs and to reason about their program's correctness. Further, many IDEs internally run static analyses and warn the user when errors arise in their code. For example, a positive answer to (i) or (ii) can clearly point the programmer to the part of the program they need to inspect in order to avoid potential bugs, whereas a positive answer to (iii) can be used by a compiler to safely move \TT{e} outside the loop body, avoiding unnecessary re-computation at each iteration in runtime.

Unfortunately, all of these questions can be reduced to fundamental problems that are proven to be undecidable. Rice's theorem~\cite{rice1953classes} states that it is undecidable to answer such questions exactly, which makes it inevitable to approximate. For instance, an approximate analysis for (i) would either answer ``no, there is definitely no use of \TT{x} before it is initialized'' or ``\emph{maybe} there is a use of \TT{x} before it is initialized.'' A major goal of static analysis is to design analyses and algorithms that achieve high precision while being tractable with respect to their application domain.

\paragraph{Industrial Applications of Static Analysis} Static analysis is widely used in the industry and can save companies great costs. The availability of such analyses, and formal verification in general, is crucial to industries that develop embedded software used to control large cyber-physical systems, the failure of which might incur a cost of hundreds of millions of dollars or even endanger human lives. To illustrate, consider avionics software used to control aircraft. In June of 1996, the first launch of the Ariane 5 rocket failed when the rocket self-destructed 40 seconds after its takeoff, a tragedy that cost \$370 million. A report later revealed that this was caused by a software error due to unsafe type conversion from a 64-bit float into a 16-bit integer. The exception raised by this run-time error was not caught, leading to undefined behavior that eventually led to the rocket's self-destruct~\cite{DBLP:journals/sigsoft/Dowson97}. As a positive example, the Astrée static analyzer~\cite{DBLP:conf/esop/CousotCFMMMR05} has been commercially used to exhaustively detect all possible runtime errors for a large class of errors such as division by zero, null-dereferencing, and deadlocks. In November 2003, Astrée was used to formally verify that the primary flight control software of the Airbus A340 aircraft contains no runtime errors. That software contains 132,000 lines of C code and was analyzed by the tool in less than 2 hours~\cite{DBLP:conf/isola/Cousot07}.

\paragraph{Data-flow Analysis} Data-flow analysis is a catch-all term for a wide and expressive variety of static program analyses that include common tasks such as reaching definitions~\cite{Collard98acomparative}, points-to and alias analysis~\cite{DBLP:conf/cgo/ShangXX12,DBLP:conf/pldi/SridharanB06,DBLP:conf/oopsla/SridharanGSB05,DBLP:conf/ecoop/XuRS09,DBLP:conf/issta/YanXR11,DBLP:conf/popl/ZhengR08}, null-pointer dereferencing~\cite{DBLP:conf/icse/NandaS09,DBLP:journals/cacm/Meyer17,DBLP:conf/atva/DasL17}, uninitialized variables~\cite{DBLP:conf/cc/NguyenIAC03} and dead code elimination~\cite{gupta1997path}, as well as several other standard frameworks, e.g.~gen-kill and bit-vector problems~\cite{knoop1993efficient,DBLP:journals/toplas/KnoopSV96,DBLP:conf/popl/Kildall73}. The common thread among data-flow analyses is that they consider certain ``data facts'' at each line of the code and then try to ascertain which data facts may/must hold at any given point~\cite{reps}. This is often achieved by a worklist algorithm that keeps discovering new data facts until it reaches a fixed point and converges to the final solution~\cite{reps,kildall1972global}. Variants of data-flow analysis are already included in most IDEs and compilers. For example, Eclipse has support for various data-flow analyses, such as unused variables and dead code elimination, both natively~\cite{eclipse} and through plugins~\cite{pessoa2012eclipse,pmd}. Data-flow analyses have also been applied in the context of compiler optimization, e.g.~for register allocation~\cite{DBLP:conf/dac/KurdahiP87} and constant propagation analysis~\cite{DBLP:conf/pldi/GroveT93,DBLP:journals/tcs/SagivRH96,DBLP:conf/sigplan/CallahanCKT86}. Additionally, they have found important use-cases in security~\cite{DBLP:conf/ccs/ChangSL08}, including in taint analysis~\cite{DBLP:conf/pldi/ArztRFBBKTOM14} and detection of SQL injection attacks~\cite{DBLP:conf/icse/GouldSD04a}. Due to their apparent importance, data-flow analyses have been widely studied by the verification, compilers, security and programming languages communities over the past five decades and are also included in program analysis frameworks such as Soot~\cite{sootifds} and WALA~\cite{wala}. 

\paragraph{Intraprocedural vs Interprocedural Analysis} Traditionally, data-flow analyses are divided into two general groups~\cite{khedker2017data}:

\begin{compactitem}
	\item \emph{Intraprocedural} approaches analyze each function/procedure of the code in isolation~\cite{DBLP:conf/popl/Kildall73,pmd}. This enables modularity and helps with efficiency, but the tradeoff is that the call-context and interactions between the different procedures are not accounted for, hence leading to relatively lower precision.
	\item In contrast, \emph{interprocedural} analyses consider the entirety of the program, i.e.~all the procedures, at the same time. They are often sensitive to call context and only focus on execution paths that respect function invocation and return rules, i.e.~when a function ends, control has to return to the correct site of the last call to that function~\cite{reps,DBLP:conf/sigplan/ChowR82}. Unsurprisingly, interprocedural analyses are much more accurate but also have higher complexity than their intraprocedural counterparts~\cite{reps2000undecidability,reps,DBLP:conf/cc/RountevKM06,DBLP:journals/pacmpl/SpathAB19}. 
\end{compactitem}

\paragraph{IFDS} We consider the standard \emph{Interprocedural Finite Distributive Subset} (IFDS) framework~\cite{reps,DBLP:journals/infsof/Reps98}. IFDS is an expressive framework that captures a large class of interprocedural data-flow analyses including the analyses enumerated above, and has been widely used in program analysis. In this framework, we assign a set $D$ of data facts to each line of the program and then apply a reduction to a variant of graph reachability with side conditions ensuring that function call and return rules are enforced. For example, in a null-pointer analysis, each data fact $d_i$ in $D$ is of the form ``the pointer $p_i$ might be null''.  See Chapter~\ref{sec:ifds} for details. Given a program with $n$ lines, the original IFDS algorithm in~\cite{reps} solves the data-flow problem \emph{for a fixed starting point} in time $O(n \cdot \vert D \vert^3).$ Due to its elegance and generality, this framework has been thoroughly studied by the community. It has been extended to various platforms and settings~\cite{DBLP:conf/pldi/ArztRFBBKTOM14,DBLP:conf/cc/NaeemLR10,DBLP:conf/pldi/BoddenTRBBM13}, notably the on-demand setting~\cite{demand} and in presence of correlated method calls~\cite{DBLP:conf/sas/RapoportLT15}, and has been implemented in standard static analysis tools~\cite{wala,sootifds}.

\paragraph{On-demand Data-flow Analysis} Due to the expensiveness of exhaustive data-flow analysis, i.e.~an analysis that considers every possible starting point, many works in the literature have turned their focus to on-demand analysis~\cite{demand,esop,DBLP:journals/acta/BabichJ78a,DBLP:conf/oopsla/SridharanGSB05,DBLP:conf/issta/YanXR11,DBLP:conf/popl/ZhengR08,DBLP:conf/popl/DuesterwaldGS95,DBLP:conf/deductive/Reps93}. In this setting, the algorithm can first run a preprocessing phase in which it collects some information about the program and produces summaries that can be used to speed up the query phase. Then, in the query phase, the algorithm is provided with a series of queries and should answer each one as efficiently as possible. Each query is of the form $(\ell_1, d_1, \ell_2, d_2)$ and asks whether it is possible to reach line $\ell_2$ of the program, with the data fact $d_2$ holding at that line, assuming that we are currently at line $\ell_1$ and data fact $d_1$ holds\footnote{Instead of single data facts $d_1$ and $d_2$, we can also use a set of data facts at each of $\ell_1$ and $\ell_2,$ but as we will see in Chapter~\ref{sec:ifds}, this does not affect the generality.}. It is also noteworthy that on-demand algorithms commonly use the information found in previous queries to handle the current query more efficiently. On-demand analyses are especially important in just-in-time compilers and their speculative optimizations~\cite{esop,DBLP:conf/cc/ChenLDHY04,DBLP:journals/taco/LinCHYJNC04,DBLP:conf/oopsla/BebenitaBFLSTV10,DBLP:journals/pacmpl/FluckigerSYGAV18}, in which having dynamic information about the current state of the program can dramatically decrease the overhead for the compiler. In addition, on-demand analyses have the following merits~(quoted from\cite{demand,DBLP:journals/infsof/Reps98}):
\begin{displayquote}
\begin{compactitem}
	\item narrowing down the focus to specific points of interest,
	\item narrowing down the focus to specific data-flow facts of interest, 
	\item reducing the work in preliminary phases, 
	\item side-stepping incremental updating problems, and \item offering on-demand analysis as a user-level operation that helps programmers with debugging.
\end{compactitem}

\end{displayquote}

\paragraph{On-demand IFDS} An on-demand variant of the IFDS algorithm was first provided in~\cite{demand}. This method has no preprocessing but memoizes the information obtained in each query to help answer future queries more efficiently. It outperforms the classical IFDS algorithm of~\cite{reps} in practice but does not provide any theoretical guarantees on the running time except that the on-demand version will never be any worse than running a new instance of the IFDS algorithm per query. Hence, the worst-case runtime on $m$ queries is $O(m \cdot n \cdot |D|^3).$ Recall that $n$ is the number of lines in the program and $|D|$ is the number of data facts at each line. Alternatively, one can push all the complexity to the preprocessing phase, running the IFDS algorithm exhaustively for each possible starting point, and then answering queries by a simple table lookup. In this case, the preprocessing will take $O(n^2 \cdot |D|^3).$ Unfortunately, none of these two variants are scalable enough to handle codebases with hundreds of thousands of lines, e.g.~standard utilities in the DaCapo benchmark suite~\cite{dacapo} such as Eclipse or Jython. In practice, software giants such as Google or Meta need algorithms that are applicable to much larger codebases, with tens or even hundreds of millions of lines.

\paragraph{Same-context On-demand IFDS} The work~\cite{esop} provides a parameterized algorithm for a special case of the on-demand IFDS problem. The key idea in~\cite{esop} is to observe that control-flow graphs of real-world programs are sparse and tree-like and that this sparsity can be exploited to obtain more efficient algorithms for \emph{same-context} IFDS queries. More specifically, the sparsity is formalized by a graph parameter called treewidth~\cite{robertson1986graph,robertson1984graph}. Intuitively speaking, treewidth measures for a graph how much it resembles a tree, i.e.~more tree-like graphs have smaller treewidth. See Chapter~\ref{sec:param} for a formal definition. It is proven that structured programs in several languages, such as C, have bounded treewidth~\cite{thorup} and there are experimental works that establish small bounds on the treewidth of control-flow graphs of real-world programs written in other languages, such as Java~\cite{DBLP:conf/alenex/GustedtMT02}, Ada~\cite{DBLP:conf/adaEurope/BurgstallerBS04} and Solidity~\cite{DBLP:conf/sac/ChatterjeeGG19}. Using these facts,~\cite{esop} provides an on-demand algorithm with a running time of $O(n \cdot |D|^3)$ for preprocessing and $O\left(\lceil \frac{|D|}{\lg n} \rceil\right)$ time per query\footnote{This algorithm uses the Word-RAM model of computation. The division by $\lg n$ is obtained by encoding $\lg n$ bits in one word.}. In practice, $|D|$ is often tiny in comparison with $n$ and hence this algorithm is considered to have linear-time preprocessing and constant-time query. Unfortunately, the algorithm in~\cite{esop} is not applicable to the general case of IFDS and can only handle \emph{same-context} queries. Specifically, the queries in~\cite{esop} provide a tuple $(\ell_1, d_1, \ell_2, d_2)$ just as in standard IFDS queries but they ask whether it is possible to reach $(\ell_2, d_2)$ from $(\ell_1, d_1)$ by an execution path that \emph{preserves the state of the stack}, i.e.~$\ell_1$ and $\ell_2$ are limited to being in the same function and the algorithm only considers execution paths in which every intermediate function call returns before reaching $\ell_2$. 

\paragraph{Our Contribution} In this work, we present a novel algorithm for the general case of on-demand IFDS analysis. Our contributions are as follows:
\begin{compactitem}

\item We identify a new sparsity parameter, namely the treedepth of the program's call graph, and use it to find a more efficient and scalable parameterized algorithm for IFDS data-flow problems. Hence, our approach exploits the sparsity of both call graphs and control-flow graphs and bounds both treedepth and treewidth. Treedepth~\cite{treedepth,DBLP:journals/siamdm/BodlaenderDJKKMT98} is a well-studied graph sparsity parameter. It intuitively measures for a graph how much it resembles a star or a shallow tree~\cite[Chapter 6]{nevsetvril2012sparsity}.

\item We provide a scalable algorithm that is not limited to same-context queries as in~\cite{esop} and is much more efficient than the classical on-demand IFDS algorithm of~\cite{demand}. Specifically, after a lightweight preprocessing that takes $O(n \cdot |D|^3 \cdot \text{treedepth})$ time, our algorithm is able to answer each query in $O(|D|^3 \cdot \text{treedepth})$. Thus, this is the first algorithm that can solve the general case of on-demand IFDS scalably and handle codebases and programs where the number of lines of code can reach hundreds of thousands or even millions.

\item We provide experimental results on the standard DaCapo benchmarks~\cite{dacapo} illustrating that:
\begin{compactitem}
	\item Our assumption of the sparsity of call graphs and low treedepth holds in practice in real-world programs; and
	\item Our approach comfortably beats the runtimes of exhaustive and on-demand IFDS algorithms~\cite{reps,demand} by two orders of magnitude.
\end{compactitem}
\end{compactitem}

\paragraph{Novelty} Our approach is novel in several directions: \begin{compactitem}
	\item Unlike previous optimizations for IFDS that only focused on control-flow graphs, we exploit the sparsity of both control-flow and call graphs.
	\item To the best of our knowledge, this is the first time that the treedepth parameter is exploited in a static analysis or program verification setting. While this parameter is well-known in the graph theory community and we argue that it is a natural candidate for formalizing the sparsity of call graphs (See Chapter~\ref{sec:param}), this is the first work that considers it in this context. 
	\item We provide the first theoretical improvements in the runtime of general on-demand data-flow analysis since~\cite{demand}, which was published in 1995. Previous improvements were either heuristics without a theoretical guarantee of improvement or only applicable to the special case of same-context queries.
	\item Our algorithm is much faster than~\cite{demand} in practice and is the first to enable on-demand interprocedural data-flow analysis for programs with hundreds of thousands or even millions of lines of code. Previously, for such large programs, the only choices were to either apply the data-flow analysis intraprocedurally, which would significantly decrease the precision, or to limit ourselves to the very special case of same-context queries~\cite{esop}.
\end{compactitem}

\paragraph{Limitation} The primary limitation of our algorithm is that it relies on the assumption of bounded treewidth for control-flow graphs and bounded treedepth for call graphs. In both cases, it is theoretically possible to generate pathological programs that have arbitrarily large width/depth: \cite{DBLP:conf/alenex/GustedtMT02} shows that it is possible to write Java programs whose control-flow graphs have any arbitrary treewidth. However, such programs are highly unrealistic, e.g.~they require a huge number of labeled nested while loops with a large nesting depth and break/continue statements that reference a while loop that is many levels above in the nesting order. Similarly, we can construct a pathological example program whose call graph has a large treedepth. Nevertheless, this is also unrealistic and real-world programs, such as those in the DaCapo benchmark suite, have both small treewidth and small treedepth, as shown in Chapter~\ref{sec:exper} and~\cite{thorup,DBLP:conf/alenex/GustedtMT02,DBLP:conf/adaEurope/BurgstallerBS04,DBLP:conf/sac/ChatterjeeGG19}.

\paragraph{Organization} In Chapter~\ref{sec:ifds}, we present the standard IFDS framework and formally define our problem. This is followed by a presentation of the graph sparsity parameters we will use, i.e.~treewidth and treedepth, in Chapter~\ref{sec:param}. Chapter \ref{sec:review} reviews related previous approaches to the problem. Our algorithm is then presented in Chapter~\ref{sec:algo}, followed by experimental results in Chapter~\ref{sec:exper} and then a conclusion in Chapter \ref{sec:conclusion}.
\chapter{The IFDS Framework} \label{sec:ifds}

In this chapter, we provide an overview of the IFDS framework following the notation and presentation of~\cite{esop,reps} and formally define the interprocedural data-flow problem considered in this work.

\paragraph{Model of Computation} Throughout this thesis, we will assume the standard RAM model of computation in which every word is of length $\Theta(\lg n)$, where $n$ denotes the size of the input. We assume that common operations, such as addition, shift and bitwise logic between a pair of words, take $O(1)$ time. Note that this has no effect on the implementation of our algorithms since most modern computers have a word size of at least $64$ and we are not aware of any possible real-world input to our problems whose size can potentially exceed $2^{64}.$ We need this assumption since we use the algorithm of~\cite{esop} as a black box. Our own contribution does not rely on the word RAM model.

\paragraph{Control-flow Graphs} In IFDS, a program with $k$ functions $f_1, f_2, \dots, f_k$ is modeled by $k$ control-flow graphs $G_1, G_2, \dots, G_k$, one for each function, as well as certain interprocedural edges that model function calls and returns. The graphs $G_i$ are standard control-flow graphs, having a dedicated \emph{start vertex} $s_i$ modeling the beginning point of $f_i$, another dedicated \emph{end vertex} $e_i$ modeling its end point, one vertex for every line of code in $f_i,$ and a directed edge from $u$ to $v,$ if line $v$ can potentially be reached right after line $u$ in some execution of the program. The only exception is that function call statements are modeled by two vertices: a \emph{call} vertex $c_l$ and a \emph{return site} vertex $r_l$. The vertex $c_l$ only has incoming edges, whereas $r_l$ only has outgoing edges. There is also an edge from $c_l$ to $r_l$, which is called a \emph{call-return-site} edge. This edge is used to pass local information, e.g.~information about the variables in $f_i$ that are unaffected by the function call, from $c_l$ to $r_l$.

\begin{figure}
	\centering
	\begin{minipage}{0.5\textwidth}
		\centering
		\lstinputlisting{figures/cfg\_code.cpp}
	\end{minipage}
	\begin{minipage}{0.45\textwidth}
		\centering
		\includegraphics[keepaspectratio,scale=0.75]{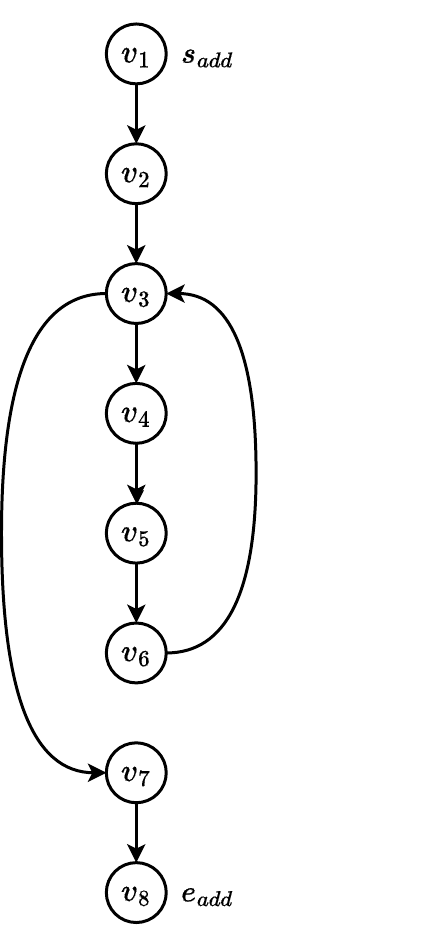}
	\end{minipage}
	\caption{To the left is a single-function C++ program and to the right is its associated control-flow graph.}
	\label{fig:cfg}
\end{figure}

\paragraph{Example} Figure~\ref{fig:cfg} shows a program consisting of one function and its corresponding control-flow graph.

\paragraph{Supergraphs} The entire program is modeled by a \emph{supergraph} $G$, consisting of all the control-flow graphs $G_i,$ as well as interprocedural edges between them. If a function call statement in $f_i,$ corresponding to vertices $c_l$ and $r_l$ in $G_i,$ calls the function $f_j$, then the supergraph contains the following interprocedural edges:
\begin{compactitem}
	\item a \emph{call-start} edge from the call vertex $c_l$ to the start vertex $s_j$ of the called function $f_j,$ and
	\item an \emph{exit-return-site} edge from the endpoint $e_j$ of the called function $f_j$ back to the return site $r_l$.
\end{compactitem}

~\\

\begin{figure}
	\centering
	\begin{minipage}{0.5\textwidth}
		\centering
\begin{lstlisting}
void g(int *&a, int *&b) {
	b = a;
}

int main() {
	int *a, *b;
	a = new int(42);
	g(a, b);
	*b = 0;
}
\end{lstlisting}
	\end{minipage}
	\begin{minipage}{0.45\textwidth}
		\centering
		\includegraphics[keepaspectratio,scale=0.75]{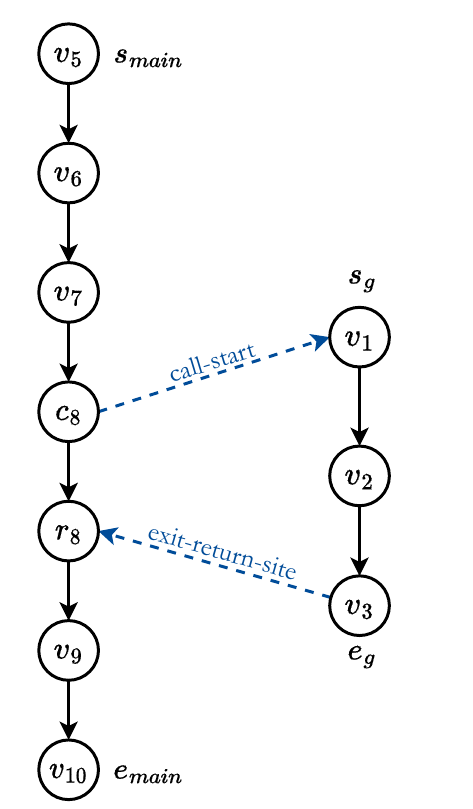}
	\end{minipage}
	\caption{To the left is a C++ program and to the right is its associated supergraph.}
	\label{fig:supergraph}
\end{figure}

\paragraph{Example} Figure~\ref{fig:supergraph} shows a program with two functions and its respective supergraph.

\paragraph{Call Graphs} Given a supergraph $G$ as above, a call graph is a directed graph $C$ whose vertices are the functions $f_1, \dots, f_k$ of the program and there is an edge from $f_i$ to $f_j$ iff there is a function call statement in $f_i$ that calls $f_j$. In other words, the call graph models the interprocedural edges in the supergraph and the supergraph can be seen as a combination of the control-flow and call graphs.

\begin{figure}
	\centering
	\begin{minipage}{0.5\textwidth}
		\centering
		\lstinputlisting{figures/call\_graph.cpp}
	\end{minipage}
	\begin{minipage}{0.45\textwidth}
		\centering
		\includegraphics[keepaspectratio,scale=0.75]{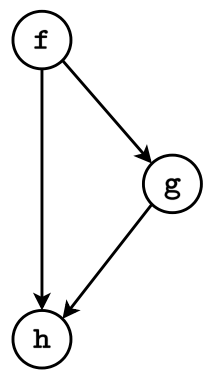}
	\end{minipage}
	\caption{To the left is a C++ program and to the right is its associated call graph.}
	\label{fig:callgraph}
\end{figure}

\paragraph{Example} Figure~\ref{fig:callgraph} shows a program consisting of 3 functions and its call graph.

\begin{figure}
	\centering
	\begin{minipage}{0.5\textwidth}
		\centering
		\lstinputlisting{figures/valid\_paths.cpp}
	\end{minipage}
	\begin{minipage}{0.45\textwidth}
		\hspace{-3em}\includegraphics[keepaspectratio,scale=0.75]{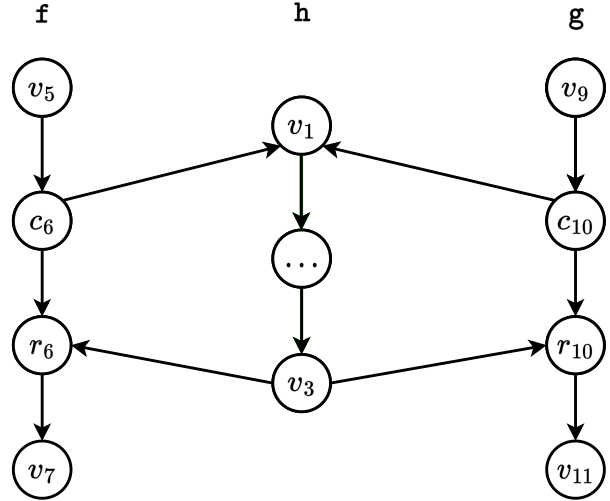}
	\end{minipage}
	\caption{To the left is a C++ program and to the right is its associated supergraph.}
	\label{fig:valid_paths}
\end{figure}

\paragraph{Valid Paths} The supergraph $G$ potentially contains invalid paths, i.e.~paths that are not realizable by an actual run of the underlying program. The IFDS framework only considers \emph{interprocedurally valid} paths in $G.$ These are the paths that respect the rules for function invocation and return. More concretely, when a function $f$ finishes execution, control should continue from the return-site vertex corresponding to the call node that called $f$. To illustrate, consider the program on the left of Figure~\ref{fig:valid_paths} and its supergraph to the right. The path $v_5\cdot c_6\cdot v_1\cdots  v_3 \cdot r_6\cdot v_7$ is a valid path since it started at $\texttt{f},$ called $\texttt{h},$ and eventually returned to $\TT{f}.$ However, the path $v_5\cdot c_6\cdot v_1\cdots  v_3 \cdot r_{10}\cdot v_{11}$ is invalid path because it returns to $\TT{g}$ rather than $\TT{h}.$ We wish to exclude the effect of such invalid paths from our analysis.

Formally, let $\Pi$ be a path in $G$ and derive from it the sub-sequence $\Pi^*$ by removing any vertex that was not a call vertex $c_l$ or a return-site vertex $r_l$. We call $\Pi$ a \emph{same-context interprocedurally valid} path if $\Pi^*$ can be derived from the non-terminal $S$ in the following grammar:
$$
	S \rightarrow~~ \epsilon~~|~~ c_l~~S~~r_l~~S.
$$
In other words, any function call in $\Pi$ that was invoked in line $c_l$ should end by returning to its corresponding return-site $r_l.$ A same-context valid path preserves the state of the function call stack. In contrast, path $\Pi$ is said to be \emph{interprocedurally valid} or just \emph{valid} if $\Pi^*$ is derived by the non-terminal $S'$ in the following grammar:
$$
S' \rightarrow~~ S~~|~~ S'~~c_l~~S .
$$
In the remainder of the thesis, we will use IVP and SCVP as abbreviations for interprocedurally valid path and same-context interprocedurally valid path respectively. An IVP has to respect the rules for returning to the right return-site vertex after the end of each function, but it does not necessarily keep the function call stack intact and is allowed to have function calls that do not necessarily end by the end of the path.

Let $u_1$ and $u_2$ be vertices in the supergraph $G$. Define $\scv(u_1,u_2)$ to be the set of all SCVPs from $u_1$ to $u_2$ by and similarly define $\valid(u_1, u_2)$ to be the set of all IVPs from $u_1$ to $u_2$. In IFDS, we only focus on valid paths and hence the problem is to compute a \emph{meet-over-all-valid-paths} solution to data-flow facts, instead of the \emph{meet-over-all-paths} approach that is usually taken in intraprocedural data-flow analysis~\cite{reps}.

\paragraph{IFDS Arena~\cite{reps}} An \emph{arena} of the IFDS data-flow analysis is a five-tuple $(G, D, \Phi, M, \meet)$ wherein:
\begin{compactitem}
	\item $G = (V, E)$ is a supergraph consisting of control-flow graphs and interprocedural edges, as illustrated above.
	\item $D$ is a finite set of \emph{data facts}. Intuitively, we would like to keep track of which subset of data facts in $D$ hold at any vertex of $G$ (line of the program).
	\item $\meet$ is the \emph{meet operator} which  is either union or intersection, i.e.~$\meet \in \{ \cup, \cap \}.$
	\item $\Phi$ is the set of \emph{flow functions}. Every function $\phi \in \Phi$ is of the form $\phi: 2^D \rightarrow 2^D$ and \emph{distributes over $\meet$}, i.e.~for every pair of subsets of data facts $D_1, D_2 \subseteq D,$ we have $\phi(D_1 \meet D_2) = \phi(D_1) \meet \phi(D_2).$
	\item $M: E \rightarrow \Phi$ is a function that maps every edge of the supergraph to a distributive flow function. Informally, $M(e)$ models the effect of executing the edge $e$ on the set of data facts. If the data facts that held before the execution of the edge $e$ are given by a subset $D' \subseteq D,$ then the data facts that hold after $e$ are $M(e)(D') \subseteq D.$  
\end{compactitem}
We can extend the function $M$ to any path $\Pi$ in $G$. Let $\Pi$ be a path consisting of the edges $e_1, e_2, \dots, e_\pi.$ We define $M(\Pi) := M(e_\pi) ~\circ~ M(e_{\pi-1}) ~\circ~ \cdots ~\circ~ M(e_1).$ Here, $\circ$ denotes function composition. According to this definition, $M(\Pi)$ models the effect that $\Pi$'s execution has on the data facts that held at the start of $\Pi$.

\paragraph{Problem Formalization} Consider an initial state $(u_1, D_1) \in V \times 2^D$ of the program, i.e.~we are at line $u_1$ of the program and we know that the data facts in $D_1$ hold. Let $u_2 \in V$ be another line, we define
$$
\mivp(u_1, D_1, u_2) ~:=~ \bigsqcap_{\Pi \in \valid(u_1, u_2) } M(\Pi)(D_1).
$$  
We simplify the notation to $\mivp(u_2),$ when the initial state is clear from the context. Our goal is to compute the $\mivp$ values. Intuitively, $\mivp$ corresponds to \emph{meet-over-all-valid-paths.} If $\meet = \cap$, then $\mivp(u_2)$ models the data facts that \emph{must} hold whenever we reach $u_2$. Conversely, if $\meet = \cup$, then $\mivp(u_2)$ corresponds to the data facts that \emph{may} hold when reaching $u_2$. The work~\cite{reps} provides an algorithm to compute $\mivp(u_2)$ for every end vertex $u_2$ in $O(n \cdot |D|^3),$ in which $n = |V|.$

\paragraph{Same-context IFDS} We can also define a same-context variant of $\mivp$ as follows:
$$
\mscvp(u_2) ~:=~ \bigsqcap_{\Pi \in \scv(u_1, u_2) } M(\Pi)(D_1).
$$ 
The intuition is similar to $\mivp,$ but in $\mscvp$ we only take into account SCVPs which preserve the function call stack's status and ignore other valid paths. The work~\cite{esop} uses parameterization by treewidth of the control-flow graphs to obtain faster algorithms for computing $\mscvp.$ However, its algorithms are limited to the same-context setting. In contrast, in this thesis, we follow the original IFDS formulation of~\cite{reps} and focus on $\mivp,$ not $\mscvp.$ Our main contribution is that we present the first theoretical improvement for computing $\mivp$ since~\cite{reps,demand}.

\paragraph{Dualization} In this work, we suppose that the meet operator is union. In other words, we focus on \emph{may} analyses. This is without loss of generality since to solve an IFDS instance with intersection as its meet operator, i.e.~a \emph{must} analysis, we can reduce it to a union instance with a simple dualization transformation. See~\cite{DBLP:conf/popl/RepsHS95} for details.

\paragraph{Data Fact Domain} In our presentation, we are assuming that there is a fixed global data fact domain $D$. In practice, the domain $D$ can differ in every function of the program. For example, in a null-pointer analysis, the data facts in each function keep track of the nullness of the pointers that are either global or local to that particular function. However, having different $D$ sets would reduce the elegance of the presentation and has no real effect on any of the algorithms. So, we follow~\cite{reps,esop} and consider a single domain $D$ in the sequel. Our implementation in Chapter~\ref{sec:exper} supports different domains for each function. 

\begin{figure}
	\centering
	\begin{minipage}{0.5\textwidth}
		\centering
\begin{lstlisting}
void g(int *&a, int *&b) {
	b = a;
}

int main() {
	int *a, *b;
	a = new int(42);
	g(a, b);
	*b = 0;
}
\end{lstlisting}
	\end{minipage}
	\begin{minipage}{0.45\textwidth}
		\hspace{-1em}\includegraphics[keepaspectratio, scale=0.75]{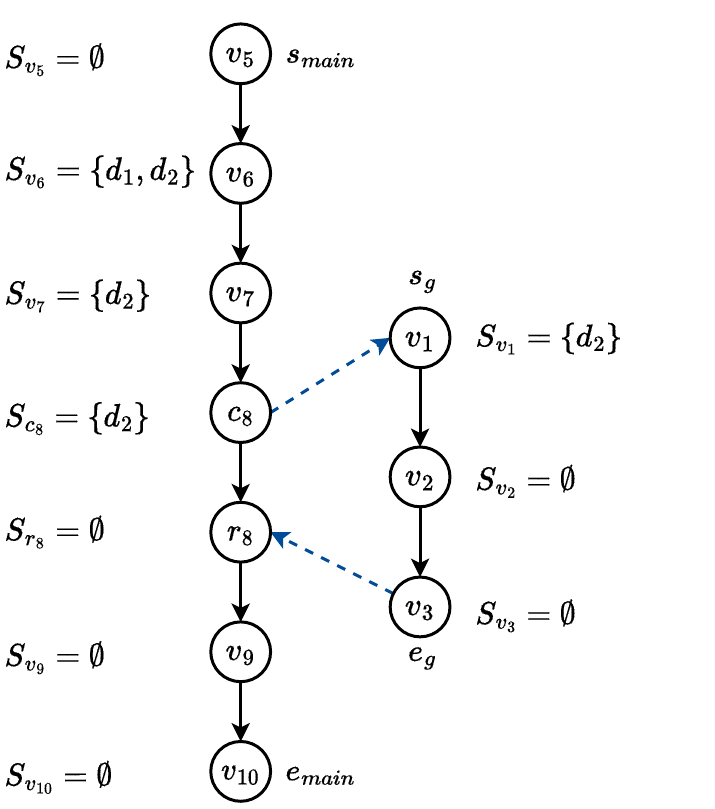}
	\end{minipage}
	\caption{To the left is a program and to the right is its associated supergraph along with values of $S_v$ for every $v$.}
	\label{fig:dataflow}
\end{figure}

\paragraph{Example} Figure~\ref{fig:dataflow} shows the same program and supergraph as in Figure~\ref{fig:supergraph}. Suppose we wish to perform a null-pointer analysis on that program. Here, our set of data facts is $D = \{d_1, d_2\}$ where $d_1$ models the fact ``the pointer \texttt{a} may be null'' and $d_2$ does the same for \texttt{b}. Starting from $v_5$, i.e.~the beginning of the \texttt{main} function, and knowing no data facts, i.e.~$D_1 = \emptyset,$ we would like to determine at every program point $v$ which variables might be null right after executing $v$. This information is captured by the value $S_v := \mivp(v_5, \emptyset, v),$ which is shown in the figure for each $v$. For instance, $S_{v_6}$ tells us that after declaring $\TT{a}$ and $\TT{b}$, any of them may be null, whereas $S_{v_{10}}$ tells us that at the end of the program's execution, neither of the variables may be null.

\paragraph{Graph Representation of Functions~\cite{reps}} Every function $\phi: 2^D \rightarrow 2^D$ that distributes over $\cup$ can be compactly represented by a relation $R_\phi \subseteq (D \cup \{\zero\}) \times (D \cup \{\zero\})$ where:
$$
R_\phi := \{ (\zero, \zero) \} ~~\cup~~ \{ (\zero, d) ~|~ d \in \phi(\emptyset) \} ~~\cup~~ \{(d_1, d_2) ~|~ d_2 \in \phi(\{d_1\}) \setminus \phi(\emptyset)\}.
$$
The intuition is that, in order to specify the union-distributive function $\phi$, it suffices to fix $\phi(\emptyset)$ and $\phi(\{d\})$ for every $d \in D.$ Then, we always have $$\phi(\{d_1, d_2, \ldots, d_r\}) = \phi(\{d_1\}) \cup \phi(\{d_2\}) \cup \cdots \cup \phi(\{d_r\}).$$ We use a new item $\zero$ to model $\phi(\emptyset),$ i.e.~$\zero~R_\phi~d \Leftrightarrow d \in \phi(\emptyset).$ To specify $\phi(\{d\}),$ we first note that $\phi(\emptyset) \subseteq \phi(\{d\}),$ so we only need to specify the elements that are in $\phi(\{d\})$ but not $\phi(\emptyset).$ These are precisely the elements that are in relation with $d$. In other words, $\phi(\{d\}) = \phi(\emptyset) \cup \{ d' ~|~ d~R_{\phi}~d' \}$. Further, we can look at $R_\phi$ as a bipartite graph $H_\phi$ with each of its parts having $D \cup \{\zero\}$ as its node set, and its edges are defined by $R_\phi$.

\begin{figure}[H]
	\begin{center}
	\includegraphics[scale=2]{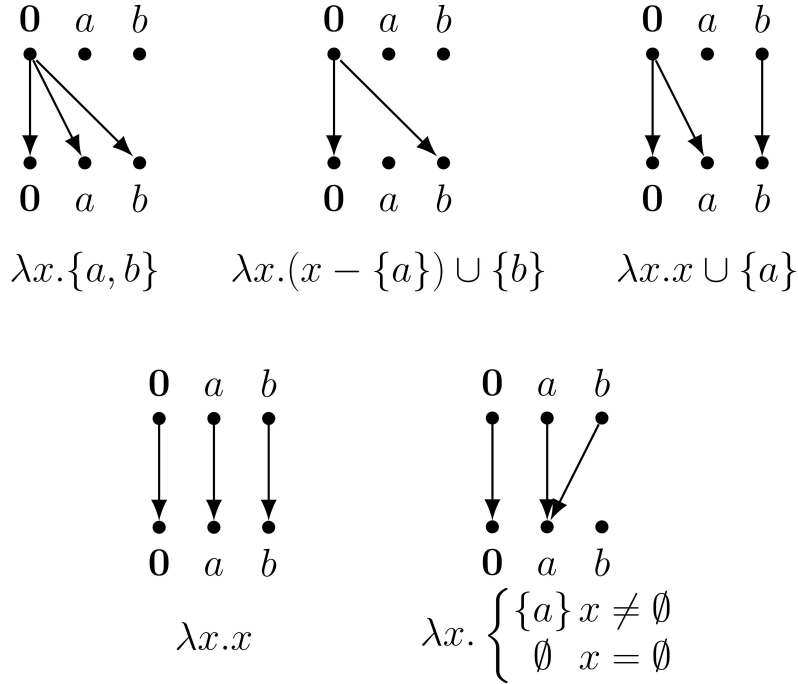}
	\end{center}
	\caption{Graph representation of union-distributive functions with $D = \{a, b\}$~\cite{esop}.}
	\label{fig:graphrep}
\end{figure}

\paragraph{Example} Figure~\ref{fig:graphrep} shows the graph representation of several union-distributive functions. 

\paragraph{Composition of Graph Representations~\cite{reps}} What makes this graph representation particularly elegant is that we can compose two functions by a simple reachability computation. Specifically, if $\phi_1$ and $\phi_2$ are distributive, then so is $\phi_2 \circ \phi_1$. By definition chasing, we can see that $R_{\phi_2 \circ \phi_1} = R_{\phi_1} ; R_{\phi_2} = \{ (d_1, d_2) ~\vert~ \exists d_3~~(d_1, d_3) \in R_{\phi_1} \wedge (d_3, d_2) \in R_{\phi_2} \}.$ Thus, to compute the graph representation $H_{\phi_2 \circ \phi_1},$ we simply merge the bottom part of $H_{\phi_1}$ with the top part of $H_{\phi_2}$ and then compute reachability from the top-most layer to the bottom-most layer. 

\begin{figure}[H]
	\begin{center}
		\includegraphics[scale=2]{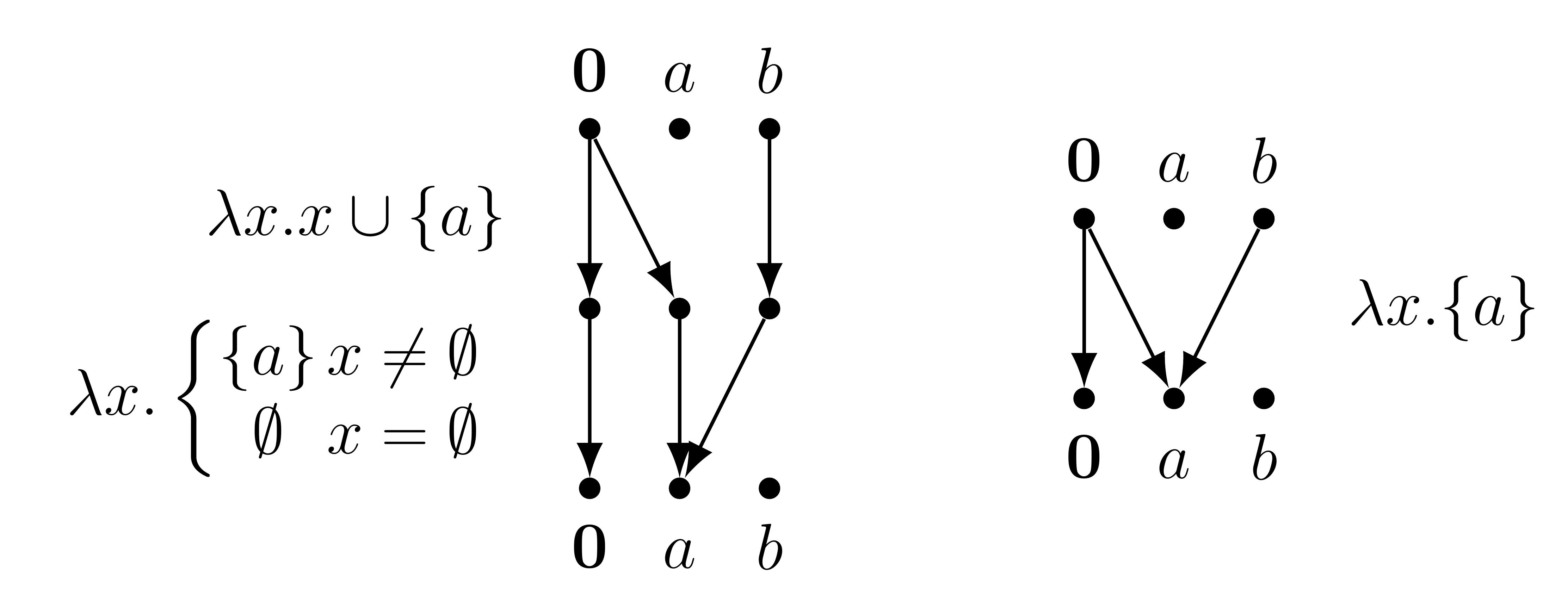}
	\end{center}
	\caption{Composing two distributive functions using reachability~\cite{esop}.}
	\label{fig:comp}
\end{figure}

\paragraph{Example} Figure~\ref{fig:comp} illustrates how the composition of two distributive functions can be obtained using their graph representations. Note that this process sometimes leads to superfluous edges. For example, since we have the edge $(\zero, a)$ in the result, the edge $(b, a)$ is not necessary. However, having it has no negative side effects, either.

\paragraph{Exploded Supergraph~\cite{reps}} Consider an IFDS arena $(G = (V, E), D, \Phi, M, \cup)$ as above and let $D^* := D\  \cup \  \{\zero\}.$ The \emph{exploded supergraph} of this arena is a directed graph $\overline{G} = (\overline{V}, \overline{E})$ in which:
\begin{compactitem}
	\item $\overline{V} = V \times D^*,$ i.e.~we take each vertex in the supergraph $G$ and copy it $|D^*|$ times; the copies correspond to the elements of $D^*.$
	\item $\overline{E} = \{ (u_1, d_1, u_2, d_2) \in \overline{V}  \times \overline{V} ~|~ (u_1, u_2) \in E~\wedge~ (d_1, d_2) \in R_{M(u_1, u_2)}\}.$ In other words, every edge between vertices $u_1$ and $u_2$ in the supergraph $G$ is now replaced by the graphic representation of its corresponding distributive flow function $M((u_1, u_2))$. 
\end{compactitem}
Naturally, we say a path $\overline{\Pi}$ in $\overline{G}$ is an IVP (SCVP) if its corresponding path $\Pi$ in $G,$ derived by extracting only the first component of vertices along $\overline{\Pi}$, is an IVP (SCVP).

\paragraph{Reduction to Reachability} We can now reformulate our problem based on reachability by IVPs in the exploded supergraph $\overline{G}.$ Consider an initial state $(u_1, D_1) \in V \times 2^D$ of the program and let $u_2 \in V$ be another line. Since the exploded supergraph contains representations of all distributive flow functions, it already encodes the changes that happen to the data facts when we execute one step of the program. Thus, it is straightforward to see that for any data fact $d_2,$ we have $d_2 \in \mivp(u_1, D_1, u_2)$ if and only if there exists a data fact $d_1 \in D_1\cup \{\zero\}$ such that the vertex $(u_2, d_2)$ in $\overline{G}$ is reachable from the vertex $(u_1, d_1)$ using an IVP~\cite{reps}. Hence, our data-flow analysis is now reduced to reachability by IVPs. Moreover, instead of computing $\mivp$ values, we can simplify our query structure so that each query consists of two nodes $(u_1, d_1)$ and $(u_2, d_2)$ in the exploded supergraph $\overline{G}$ and the query's answer is whether there exists an IVP from $(u_1, d_1)$ to $(u_2, d_2)$.

\begin{figure}
	\centering
	\begin{minipage}{0.5\textwidth}
		\centering
\begin{lstlisting}
void g(int *&a, int *&b) {
	b = a;
}

int main() {
	int *a, *b;
	a = new int(42);
	g(a, b);
	*b = 0;
}
\end{lstlisting}
	\end{minipage}
	\begin{minipage}{0.45\textwidth}
		\centering
		\includegraphics[width=0.9\textwidth]{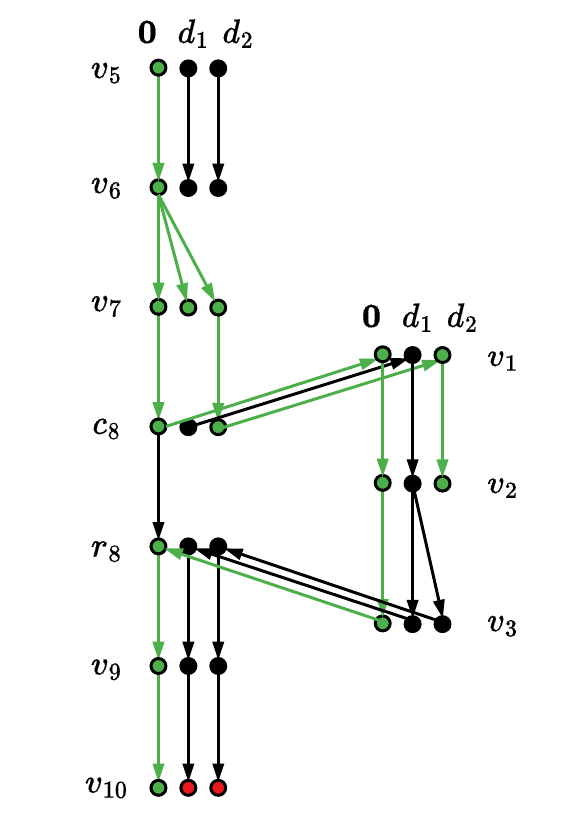}
	\end{minipage}
	\caption{To the left is a program and to the right is its associated exploded supergraph.}
	\label{fig:exploded}
\end{figure}

\paragraph{Example} Figure~\ref{fig:exploded} shows again the same program as in Figure~\ref{fig:supergraph}, together with its exploded supergraph for null-pointer analysis. $d_1$ and $d_2$ have the same meaning as in Figure~\ref{fig:dataflow}. Suppose we wish to compute $\mivp(v_5, \emptyset, v_{10}),$ i.e. which variables may be null at the end of the program assuming we start from $\TT{main},$ and that initially none of the data facts hold. Using a reachability analysis on the exploded supergraph, we can identify all vertices that can be reached by a valid path from $(v_5, \zero)$ (green) and conclude that neither \texttt{a} nor \texttt{b} may be null at the end of the program, which is consistent with the answer $S_{v_{10}} = \emptyset$ of Figure~\ref{fig:dataflow}.

\paragraph{On-demand Analysis} As mentioned in Chapter~\ref{sec:intro}, we focus on on-demand analysis and distinguish between a preprocessing phase in which the algorithm can perform a lightweight pass over the input and a query phase in which the algorithm has to respond to a large number of queries. The queries appear in a stream and the algorithm has to handle each query as fast as possible.

\paragraph{Format of Queries} Based on the discussion above, each query is of the form $(u_1, d_1, u_2, d_2) \in V \times D^* \times V \times D^*$. We define the predicates $\qry(u_1, d_1, u_2, v_2)$ and $\scqry(u_1, d_1, u_2, v_2)$ to be true if there exists an IVP, respectively SCVP, from $(u_1, d_1)$ to $(u_2, d_2)$ in $\overline{G}$ and false otherwise. The algorithm should report the truth value of $\qry(u_1, d_1, u_2, v_2)$.

\paragraph{Bounded Bandwidth Assumption} Following previous works such as~\cite{reps,esop,demand}, we assume that function calls and returns have bouned ``bandwidth''. More concretely, we assume there exists a bounded constant $\beta$ such that for every interprocedural call-start or exit-return-site edge $e$ in our supergraph $G$, the degree of each vertex in the graph representation $H_{M(e)}$ is at most $\beta.$ This is a standard assumption made in IFDS and all of its extensions. Intuitively, it captures the idea that for a function $f_i$ calling $f_j$, each parameter in $f_j$ depends on only a small number of variables in the call site line $c$ of $f_i$, and conversely, that the return value of $f_j$ depends on only a small number of variables at its last line.

\chapter{Treewidth and Treedepth} \label{sec:param}
In this chapter, we provide a short overview of the concepts of treewidth and treedepth. Treewidth and treedepth are both graph sparsity parameters and we will use them in our algorithms in the next two chapters to formalize the sparsity of control-flow graphs and call graphs, respectively.

\paragraph{Tree Decompositions~\cite{robertson1986graph,robertson1984graph,DBLP:journals/actaC/Bodlaender93}} Given an undirected graph $G = (V, E),$ a \emph{tree decomposition} of $G$ is a rooted tree $T = (\bags, E_T)$ such that:
\begin{compactenum}[(i)]
	\item Every node $b \in \bags$ of the tree $T$ has a corresponding subset $V_b \subseteq V$ of vertices of $G$. To avoid confusion, we refer to a node in $T$ as \emph{``bag"} and use the term \emph{``vertex"} only for vertices of $G$. This is natural since each bag $b$ has a subset $V_b$ of vertices.
	\item $\bigcup_{b \in \bags} V_b = V$. In other words, every vertex appears in some bag.
	\item $\forall u, v \in V, \{u, v\} \in E \implies \exists b \in \bags~~ \{u, v\} \subseteq V_b.$ That is, for every edge, there is a bag that contains both of its endpoints.
	\item For every vertex $v \in V$, the set of bags $b\in \bags$ such that $v\in V_b$ forms a connected subtree of $T$. Equivalently, if $b$ is on the unique path from $b'$ to $b''$ in $T,$ then $V_b \supseteq V_{b'} \cap V_{b''}.$
\end{compactenum}
When talking about tree decompositions of directed graphs, we simply ignore the orientation of the edges and consider decompositions of the underlying undirected graph. Intuitively, a tree decomposition covers the graph $G$ by a number of bags\footnote{The bags do not have to be disjoint.} that are connected to each other in a tree-like manner. If the bags are small, we are then able to perform dynamic programming on $G$ in a very similar manner to trees~\cite{DBLP:conf/icalp/Bodlaender88,DBLP:conf/popl/ChatterjeeGIP16,DBLP:journals/dam/GoharshadyHM16,DBLP:journals/ress/GoharshadyM20,DBLP:conf/blockchain2/MeybodiGHS22}. This is the motivation behind the following definition.

\paragraph{Treewidth~\cite{robertson1986graph}} The \emph{width} of a tree decomposition is defined as the size of its largest bag minus 1, i.e.~$w(T) := \max_{b \in \bags} |V_b| - 1.$ The \emph{treewidth} of a graph $G$ is the smallest width amongst all of its tree decompositions. Informally speaking, treewidth is a measure of tree-likeness. Only trees and forests have a treewidth of $1,$ and, if a graph $G$ has treewidth $k$, then it can be decomposed into bags of size at most $k+1$ that are connected to each other in a tree-like manner. 

\begin{figure}
	\begin{center}
		\includegraphics[keepaspectratio,scale=0.75]{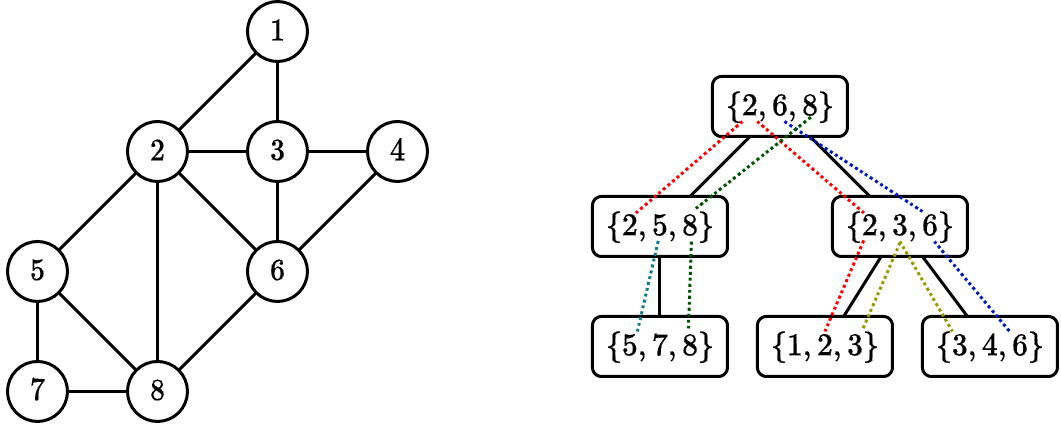}
	\end{center}
	\caption{To the left is a graph $G$ and to the right is a tree decomposition of it.}
	\label{fig:tw}
\end{figure}

\paragraph{Example} Figure~\ref{fig:tw} shows a graph $G$ on the left and a tree decomposition of width $2$ for $G$ on the right. In the tree decomposition, we have highlighted the connected subtree of each vertex by dotted lines. This tree decomposition is optimal and hence the treewidth of $G$ is $2.$

\paragraph{Computing Treewidth} In general, it is NP-hard to compute the treewidth of a given graph. However, for any fixed constant $k$, there is a linear-time algorithm that decides whether the graph has treewidth at most $k$ and, if so, also computes an optimal tree decomposition~\cite{DBLP:conf/stoc/Bodlaender93}. As such, most treewidth-based algorithms assume that an optimal tree decomposition is given as part of the input.

\paragraph{Treewidth of Control-flow Graphs} In~\cite{thorup}, it was shown that the control-flow graphs of \texttt{goto}-free programs in a number of languages such as C and Pascal have a treewidth of at most $7$. Moreover,~\cite{thorup} also provides a linear-time algorithm that, while not necessarily optimal, always outputs a tree decomposition of width at most $7$ for the control-flow graph of programs in these languages by a single pass over the parse tree of the program.
Alternatively, one can use the algorithm of~\cite{DBLP:conf/stoc/Bodlaender93} to ensure that an optimal decomposition is used at all times. The theoretical bound of~\cite{thorup} does not apply to Java, but the work~\cite{DBLP:conf/alenex/GustedtMT02} showed that the treewidth of control-flow graphs in real-world Java programs is also bounded. This bounded-treewidth property has been used in a variety of static analysis and compiler optimization tasks to speed up the underlying algorithms~\cite{DBLP:journals/toplas/ChatterjeeIGP18,DBLP:conf/cav/Obdrzalek03,DBLP:journals/siglog/Aiswarya22,DBLP:conf/fsttcs/ChatterjeeIP21,DBLP:conf/lpar/FerraraPV05,DBLP:conf/cav/ChatterjeeIP15,DBLP:journals/toplas/ChatterjeeGGIP19,DBLP:journals/pacmpl/ChatterjeeGOP19,DBLP:conf/atva/AsadiCGMP20,DBLP:phd/hal/Goharshady20,DBLP:conf/pldi/AhmadiDGP22}. Nevertheless, one can theoretically construct pathological examples with high treewidth.

\paragraph{Separators} The treewidth-based algorithm presented in Chapter~\ref{sec:review} depends on identifying certain cuts in the graph $G = (V, E)$. Let $P, Q \subseteq V$ be sets of vertices. We say the pair $(P, Q)$ is a \emph{separation} of $G$ if (i)~$P \cup Q = V,$ and (ii)~there is no edge in $G$ that connects $P \setminus Q$ to $Q \setminus P.$ In this case, we say that $P \cap Q$ is a \emph{separator}.

\paragraph{Cut Property~\cite{cygan2015parameterized}} Let $T=(\bags, E_T)$ be a tree decomposition for $G= (V, E)$ and $e = \{b, b'\} \in E_T$ an arbitrary edge of the tree. By deleting $e,$ $T$ will break into two connected subtrees $T^b$ and $T^{b'},$ containing $b$ and $b',$ respectively. Define $P := \bigcup_{c \in T^b} V_c$ and $Q := \bigcup_{c \in T^{b'}} V_c.$ Then, $(P, Q)$ is a separation of $G$ and its separator is $P \cap Q = V_b \cap V_{b'}.$

\begin{figure}
	\begin{center}
		\includegraphics[keepaspectratio,scale=0.75]{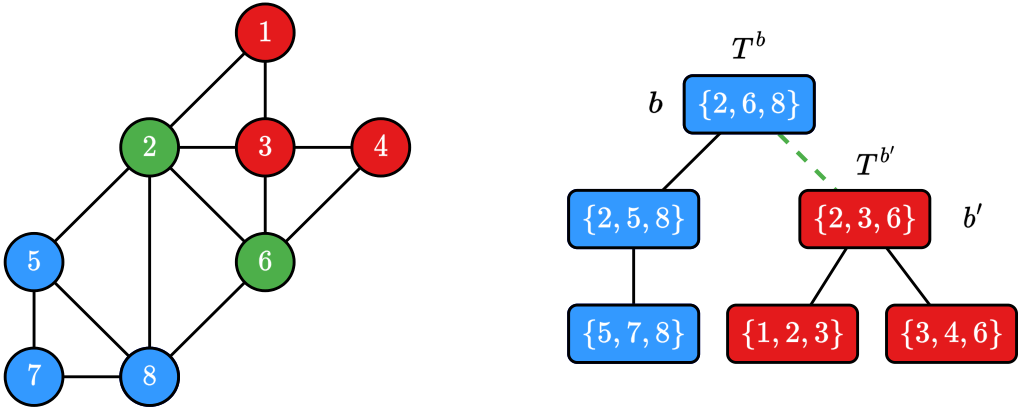}
		\caption{The cut property in tree decompositions.}
		\label{fig:cut}
	\end{center}
\end{figure}

\paragraph{Example} Figure~\ref{fig:cut} shows what happens if we cut the edge between the bags $b$ with $V_b = \{2, 6, 8\}$ and $b'$ with $V_{b'} = \{2, 3, 6\}$ in the tree decomposition of Figure~\ref{fig:tw}. The tree breaks into two parts $T^b$ (shown in blue) and $T^{b'}$ (shown in red). The only vertices that appear on both sides are $V_b \cap V_{b'} = \{2, 6\}$. These vertices are a separator in the original graph (shown in green) that separate red and blue vertices, corresponding to the red and blue parts of the tree. In other words, any path from a red vertex in $G$ to a blue vertex has to pass through one of the green vertices.

\paragraph{Balancing Tree Decompositions} The runtime of our algorithm in Chapter~\ref{sec:algo} depends on the height of the tree decomposition. Fortunately,~\cite{bodlaender1998parallel} provides a linear-time algorithm that, given a graph $G$ and a tree decomposition of constant width $t$, produces a binary tree decomposition of height $O(\lg n)$ and width $O(t).$ Combining this with the algorithms of~\cite{thorup} and~\cite{DBLP:conf/stoc/Bodlaender93} for computing low-width tree decompositions allows us to assume that we are always given a balanced and binary tree decomposition of bounded width for each one of our control-flow graphs as part of our IFDS input.

We now switch our focus to the second parameter that appears in our algorithms, namely treedepth. 

\paragraph{Partial Order Trees~\cite{treedepth}} Let $G = (V, E)$ be an undirected connected graph. A \emph{partial order tree} (POT)\footnote{The name \emph{partial order tree} is not standard in this context, but we use it throughout this work since it provides a good intuition about the nature of $T$. Usually, the term ``treedepth decomposition'' is used instead.} over $G$ is a rooted tree $T=(V, E_T)$ on the same set of vertices as $G$ that additionally satisfies the following property:
\begin{compactitem}
	\item For every edge $\{u, v\} \in E$ of $G$, either $u$ is an ancestor of $v$ in $T$ or $v$ is ancestor of $u$ in $T.$ 
\end{compactitem}
The intuition is quite straightforward: $T$ defines a partial order $\prec_T$ over the vertices $V$ in which every element $u$ is assumed to be smaller than its parent $p_u,$ i.e.~$u \prec_T p_u$. For $T$ to be a valid POT, every pair of vertices that are connected by an edge in $G$ should be comparable in $\prec_T.$ If $G$ is not connected, then we will have a partial order forest, consisting of a partial order tree for each connected component of $G$. With a slight abuse of notation, we call this a POT, too.

\begin{figure}
	\begin{center}
		\includegraphics[keepaspectratio,scale=0.75]{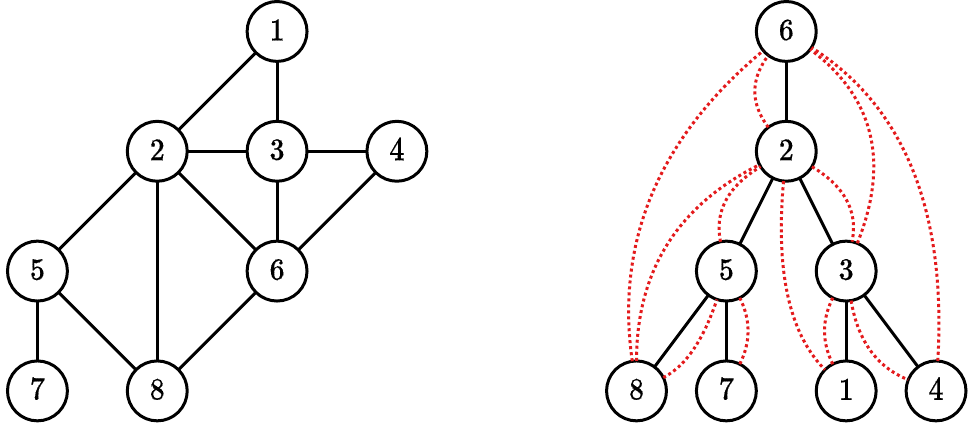}
	\end{center}
	\caption{To the left is a graph $G$ and to the right is a POT for $G$ of depth 4.}
	\label{fig:td}
\end{figure}

\paragraph{Example} Figure~\ref{fig:td} shows a graph $G$ on the left together with a POT $T$ of depth $4$ for $G$ on the right. In the POT, the edges of the original graph $G$ are shown by dotted red lines. Every edge of $G$ goes from a node in $T$ to one of its ancestors.

\paragraph{Treedepth~\cite{treedepth}} The treedepth of an undirected graph $G$ is the smallest depth among all POTs of $G.$ 

\begin{figure}
	\begin{center}
		\includegraphics[keepaspectratio,scale=0.75]{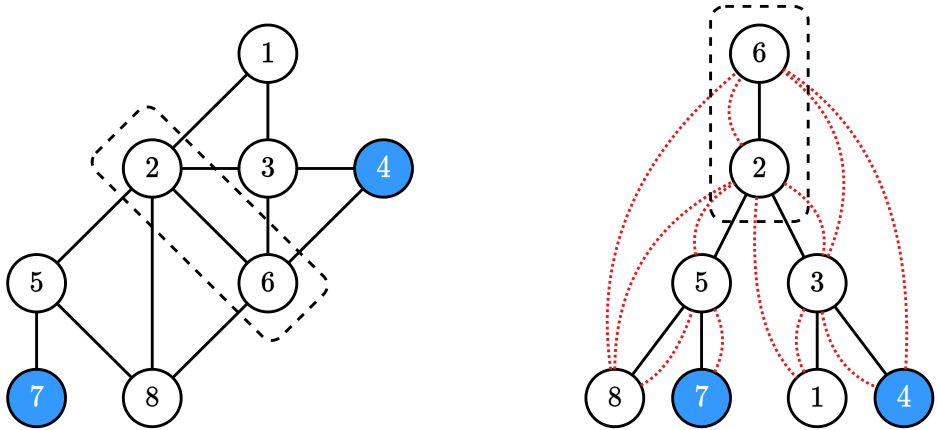}
	\end{center}
	\caption{The path property of POTs applied to nodes 4 and 7.}
	\label{fig:tdpath}
\end{figure}

\paragraph{Path Property of POTs~\cite{treedepth}} Let $T=(V, E_T)$ be a POT for a graph $G=(V, E)$ and $u$ and $v$ two vertices in $V$. Define $A_u$ as the set of ancestors of $u$ in $T$ and define $A_v$ similarly. Let $A := A_u \cap A_v$ be the set of common ancestors of $u$ and $v$. Then, any path that goes from $u$ to $v$ in the graph $G$ has to intersect $A,$ i.e.~it has to go through a common ancestor.
\newpage
\paragraph{Example} Figure~\ref{fig:tdpath} illustrates the path property where $u$ and $v$ are taken to be the nodes 4 and 7. The set of common ancestors $A = \{2, 6\}$ is enclosed by the dashed rectangle in $T$ as well as $G.$ Every path in $G$ connecting nodes $4$ and $7$ must go through either 2 or 6.

\paragraph{Sparsity Assumption} In the sequel, our algorithm is going to assume that call graphs of real-world programs have small treedepth. We establish this experimentally in Chapter~\ref{sec:exper}. However, there is also a natural reason why this assumption is likely to hold in practice. Consider the functions in a program. It is natural to assume that they were developed in a chronological order, starting with base (phase 1) functions, and then each phase of the project used the functions developed in the previous phases as libraries. Thus, the call graph can be partitioned into a small number of layers based on the development phase of each function. Moreover, each function typically calls only a small number of previous functions. So, an ordering based on the development phase is likely to give us a POT with a small depth. The depth would typically depend on the number of phases and the degree of each function in the call graph, but these are both small parameters in practice. 

\paragraph{Pathological Example} It is possible in theory to write a program whose call graph has an arbitrarily large treedepth. However, such a program is not realistic. Suppose that we want a program with treedepth $n.$ We can create $n$ functions $f_1, f_2, \ldots, f_n$ and then ensure that each function $f_i$ calls every other function $f_j~~(j \neq i).$ In this strange program, our call graph will simply be a complete graph on $n$ vertices. Since every two vertices in this graph have to be comparable, its POT will be a path with depth $n.$ So, its treedepth is $\Theta(n)$.

\paragraph{Computing Treedepth} As in the case of treewidth, it is NP-hard to compute the treedepth of a given graph~\cite{pothen1988complexity}. However, for any fixed constant $k,$ there is a linear-time algorithm that decides whether a given graph has treedepth at most $k$ and, if so, produces an optimal POT \cite{DBLP:journals/corr/abs-2205-02656}. Thus, in the sequel, we assume that all inputs include a POT of the call graph with bounded depth. 
\chapter{Previous Approaches} \label{sec:review}

In this chapter, we present various important ideas from two existing approaches to tackle on-demand IFDS problems, on top of which our algorithm in Chapter \ref{sec:algo} builds. The first part is due to the authors of the IFDS model~\cite{reps} and involves no exploitation of the graph parameters of Chapter~\ref{sec:param}. The second part is based on the work~\cite{esop}, which uses parameterization of control-flow graphs by treewidth to obtain an efficient algorithm to answer \emph{same-context} queries. In Chapter~\ref{sec:algo} we will extend these ideas to handle arbitrary interprocedural queries efficiently.

Throughout this chapter, we fix an IFDS arena $(G, D, \Phi, M, \cup)$ given by an exploded supergraph $\overline{G}$. Recall that our program has $k$ functions $F := \{f_1, \dots, f_k\}$ and $f_i$ has a control-flow graph $G_i = (V_i, E_i)$. We define $fg: V \rightarrow F$ as a function that maps each supergraph node to the program function it lies in. We assume that every $G_i$ comes with a balanced binary tree decomposition $T_i = (\bags_i, E_{T_i})$ of width at most $k_1.$ We also assume that a POT of depth $k_2$ over the call graph is given as part of the input. All these assumptions are without loss of generality since the tree decompositions and POT can be computed in linear time using the algorithms mentioned in Chapter~\ref{sec:param}. Finally, when discussing running times, we will use $D$ rather than $|D|$ to denote the size of the data facts domain.

\paragraph{Function Summaries} A core idea in~\cite{reps} is the use of \emph{function summaries} which summarize the aggregate effect of executing a function from start to exit with potential calls to other functions occurring during the execution. More formally, for a function $f_i$, we define the summary of $f_i$ to be $\summ(f_i)\subseteq D^* \times D^*$ satisfying
\[
(d_1, d_2) \in \summ(f_i) \iff \scqry(s_i, d_1, e_i, d_2).
\]
In other words, $(d_1, d_2)\in\summ(f_i)$ tells us that there is an SCVP that starts executing $f_i$ where $d_1$ holds, and exits the function with $d_2$ holding. Thus, $\chi(f_i)$ gives us complete information that summarizes $f_i$'s input/output behavior. Once summaries are computed for every function, this gives us the means to reduce the problem to standard graph reachability, as we shall show below.

\paragraph{Computing Summaries~\cite{reps}} We will sketch a polynomial-time worklist algorithm that computes summaries for all functions and outputs a new graph $\hat{G}$ such that standard reachability queries on $\hat{G}$ correspond to the answers to IVP/SCVP queries. The algorithm is essentially the same as~\cite{reps} except that they find function summaries $(d_1, d_2)\in \summ(f_i)$ only if $(s_i, d_1)$ is reachable from a fixed set of starting nodes, whereas we disregard that condition. This tweak is natural since we consider a setting where the query's starting point can be arbitrary. The variant of~\cite{reps}'s algorithm presented here is also used in~\cite{esop}.

We initially we set $\hat{G} := \overline{G}$. Denote the edges of $\hat{G}$ with $\hat{E}$. The main idea is to maintain for every function $f_i$ a list of \emph{partial summaries} that correspond to prefixes of SCVPs in $\overline{G}$ that ends at some node in $f_i$ that is not necessarily $e_i$. We keep extending those partial summaries using the intraprocedural edges of $\overline{G}$, and when a summary $(d_1, d_2)\in \summ(f_i)$ is discovered, we propagate that information to all the nodes calling $f_i$ by adding \emph{summary edges} to $\hat{G}$, which can be seen as shortcuts for SCVPs in the caller function. Those summary edges can further help a caller function discover more of its summaries. For every $f_i\in F,$ let $L_i\subseteq D^*\times V_i\times D^*$ be the list of partial summaries of $f_i$, where $(d_1, u_2, d_2)\in L_i \implies \scqry(s_i, d_1, u_2, d_2)$. 

The pseudocode of Algorithm \ref{alg:summary} shows how to compute function summaries. Initially, we have $L_i := \{(d, s_i, d) \; | \; d \in D^*\}$. Let $Q$ be a queue of partial summaries, to which we add all the initial partial summaries $L_1, L_2, \dots, L_k$, and let $Pr$ be a set that contains all the processed partial summaries, which is initially empty. These steps are done in lines 1-8. The algorithm proceeds in iterations. In every iteration, it takes a partial summary out of $Q$ and processes it. When $Q$ is empty, the algorithm returns $\hat{G}$, which is nothing more than $\overline{G}$ augmented with summary edges, and then the algorithm terminates.
Suppose we are processing $(d_1, u_2, d_2)$ and suppose $fg(u_2) = f_i$. We have two cases:
\begin{itemize}
\item $u_2 \neq e_i$ (lines 13-18): In this case, we go through all intraprocedural and summary edges originating from $(u_2, d_2)$ in $\hat{G}$. For such an edge $((u_2, d_2), (u_3, d_3))\in \hat{E}$, we know that $(d_1, u_3, d_3)$ is also a valid partial summary. We first check if $(d_1, u_3, d_3)\in Pr$, i.e.~whether it has been processed before, and if not, we add it to $L_i$, $Q$, and $Pr$.
\item $u_2 = e_i$ (lines 19-30): In this case, $(d_1, e_i, d_2)$ already forms a summary $(d_1, d_2)\in \summ(f_i)$, which implies there is an SCVP $P$ in $f_i$ from $(s_i, d_1)$ to $(e_i, d_2)$. This means that for every call node $c$ calling $f_i$ and its corresponding return-site node $r$, we can use our knowledge that $(d_1, d_2)\in \summ(f_i)$ to potentially discover more summaries in $fg(c)$. Suppose $fg(c) = f_j$. For every $d_3, d_4\in D^*$ with $((c, d_3), (s_i, d_1))\in \overline{E}$ and $((e_i, d_2), (r, d_4))\in\overline{E}$, observe that the path $$((c, d_3), (s_i, d_2))\cdot P \cdot ((e_i, d_2), (r, d_4))$$ is an SCVP in $f_j$ and therefore we add to $\hat{G}$ the \emph{summary edge} $((c, d_3), (r, d_4))$ (unless it was already added before). Further, we loop through every $d_5$ with $(d_5, c, d_3)\in L_j$, and add the extended partial summary $(d_5, r, d_4)$ to $L_j$, $Q$, and $Pr$ provided that it has not been processed before. Note that it is possible that we have a case where $(d_5, c, d_3)$ is not in $L_j$ but is added to $L_j$ in a later iteration, in which case the partial summary will be extended by the previous case through the summary edge we added (line 14).
\end{itemize}

After $Q$ becomes empty, we return $\hat{G}$. Using an inductive argument, it is not difficult to prove that the algorithm correctly finds all summaries, i.e.~upon termination we have $\summ(f_i) = \{(d_1, d_2)~|~(d_1, e_i, d_2)\in L_i\}$ for all $i$, and hence all summary edges are included in $\hat{G}$. Further, we can show that the running time is $O(n\cdot D^3)$ where $n$ is the number of lines in the program. Finally, note that the size of $\hat{G}$ is bounded by $O(n\cdot D^2)$ because for every edge $(u_1, u_2)\in G$ in the supergraph, there can be at most $O(D^2)$ edges in $\hat{G}$, which happens if we have the edges $((u_1, d_1), (u_2, d_2))$ present in $\hat{G}$ for all $d_1, d_2\in D^*$.
\begin{algorithm}
\caption{Computing function summaries.}\label{alg:summary}
\DontPrintSemicolon
$\hat{G} \gets \overline{G}$.\;
$L_1 \gets \emptyset, L_2\gets \emptyset, \dots, L_k \gets \emptyset$.\;
$Q\gets \emptyset$.\;
$Pr\gets \emptyset$.\;
\ForEach{$i \in 1..k$}{
	\ForEach{$d\in D^*$}{
		$L_i \gets L_i\cup (d, s_i, d)$.\;
		$Q \gets Q\cup (d, s_i, d)$.\;
	}
}
\While{$Q\neq \emptyset$}{
	Pick an element $(d_1, u_2, d_2)$ from $Q$.\;
	Remove $(d_1, u_2, d_2)$ from $Q$.\;
	Suppose $fg(u_2) = f_i$.\;
	\uIf{$u_2\neq e_i$}{
		\ForEach{$(u_3, d_3)$ s.t. $((u_2, d_2), (u_3, d_3))\in \hat{E}$ and $((u_2, d_2), (u_3, d_3))$ is an intraprocedural or a summary edge}{
			\If{$(d_1, u_3, d_3)\notin Pr$}{
				$L_i \gets L_i \cup (d_1, u_3, d_3)$.\;
				$Q \gets Q \cup (d_1, u_3, d_3)$.\;
				$Pr \gets Pr \cup (d_1, u_3, d_3)$.\;
			}
		} 
	}
	\Else{
		\ForEach{call node $c$ that calls $f_i$}{
			$r\gets$ the return-site associated with $c$.\;
			Suppose $fg(c) = f_j$.\;
			\ForEach{$d_3, d_4\in D^*$ s.t. $((c, d_3), (s_i, d_1))\in \overline{E}$ and $((e_i, d_2), (r, d_4))\in\overline{E}$}{
				\If{$((c, d_3), (r, d_4))\notin \hat{E}$}{
					$\hat{E}\gets \hat{E}\cup ((c, d_3), (r, d_4))$. \tcp{Summary edge.}
					\ForEach{$d_5\in D^*$ s.t. $(d_5, c, d_3)\in L_j$}{
						\If{$(d_5, r, d_4)\notin Pr$}{
							$L_j \gets L_j \cup (d_5, r, d_4)$.\;
							$Q\gets Q \cup (d_5, r, d_4)$.\;
							$Pr \gets Pr \cup (d_5, r, d_4)$.\;
						}
					}
				}
			}
		}
		
	}
}
Return $\hat{G}$.\;
\end{algorithm}

\begin{figure}[H]
	\centering
	\begin{minipage}{0.45\textwidth}
			\centering
			\includegraphics[width=0.9\textwidth]{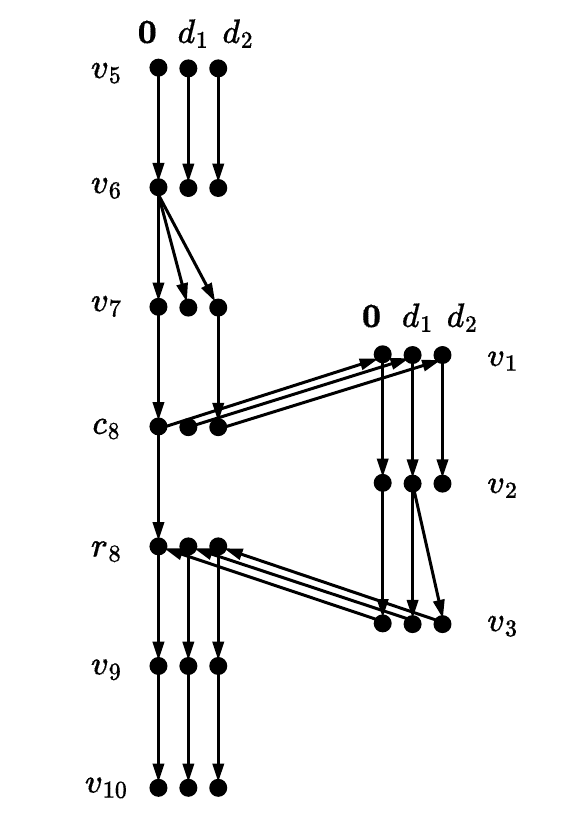}
	\end{minipage}
	\begin{minipage}{0.45\textwidth}
		\centering
		\includegraphics[width=0.9\textwidth]{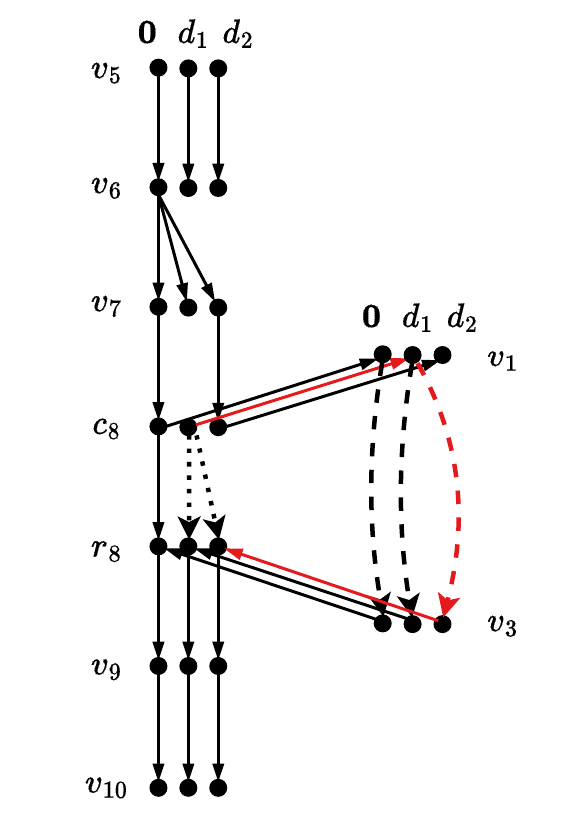}
	\end{minipage}
	\caption{To the left is an exploded supergraph and to the right is its state after computing $\summ(\texttt{g})$.}
	\label{fig:summaries}
\end{figure}

\paragraph{Example} Figure \ref{fig:summaries} shows a possible state of $\hat{G}$ while running the algorithm above on the exploded supergraph of Figure \ref{fig:exploded} and after computing $\summ(\texttt{g})$, whose pairs are denoted by the dashed edges. The dotted edges are summary edges which are added between call-return-site pair $(c_8, r_8)$ in response to discovering summaries of $\texttt{g}$. The original edges of $\texttt{g}$ are omitted on the right to avoid clutter. Initially, we have $L_\texttt{f} := \{(\zero, v_5, \zero), (d_1, v_5, d_1),  (d_2, v_5, d_2)\}$, $L_\texttt{g} := \{(\zero, v_1, \zero), (d_1, v_1, d_1),  (d_2, v_1, d_2)\}$, and $Q := L_\texttt{f} \cup L_\texttt{g}$. Suppose the queue processes partial summaries in $\texttt{g}$ first. When $(d_1, v_1, d_1)$ is processed, the intraprocedural edge $((v_1, d_1), (v_2, d_1))$ is examined and $(d_1, v_2, d_1)$ is added to $Q$. When $(d_1, v_2, d_1)$ is processed, the outgoing edges of $(v_2, d_1)$ are examined, subsequently adding the partial summaries $(d_1, v_3, d_1)$ and $(d_1, v_3, d_2)$ to $Q$. Both of these are summaries of $\texttt{g}$, and as a result, we add a summary edge $((c_8, d_1), (r_8, d_2))$ corresponding to a shortcut for the path highlighted in red. We similarly add a summary edge $((c_8, d_1), (r_8, d_1))$.
\paragraph{Reduction to Basic Reachability} We now use the output graph $\hat{G}$ of the above algorithm and further construct two graphs $\hat{G}_{\scv}$ and  $\hat{G}_{\valid}$ as follows: $\hat{G}_{\scv}$ is obtained from $\hat{G}$ after removing all call-start and exit-return-site edges, i.e.~all interprocedural edges, whereas $\hat{G}_{\valid}$ is obtained from $\hat{G}$ after only removing all exit-return-site edges and leaving the call-start edges intact. For an arbitrary graph $Gr = (V_{Gr}, E_{Gr})$ and two nodes $u, v \in V_{Gr}$, denote the reachability from $u$ to $v$ in $Gr$ by $u \leadsto_{Gr} v$. Note that this notation is concerned with standard graph reachability with no additional constraints. We will now show a correspondence between the relation $\leadsto_{\hat{G}_{\scv}}$ and SCVPs~\cite{esop}, and another more general correspondence relating $\leadsto_{\hat{G}_{\valid}}$ and IVPs. Observe the following:
\begin{itemize}
\item In $\hat{G}_{\scv}$, there can only be a path $P$ from $(u_1, d_1)$ to $(u_2, d_2)$ if we have $fg(u_1) = fg(u_2)$, because we removed all interprocedural edges. Moreover, every edge in $P$ is either an intraprocedural edge of $\overline{G}$ or a summary edge, each of which corresponds to an SCVP and hence we conclude that $$(u_1, d_1) \leadsto_{\hat{G}_{\scv}} (u_2, d_2)\implies \scqry(u_1, d_1, u_2, d_2).$$ By the correctness of Algorithm~\ref{alg:summary}, we know that it finds all possible summary edges. Thus, if there is an SCVP from $(u_1, d_1)$ to $(u_2, d_2)$, then $\hat{G}_{\scv}$ will have all necessary summary edges to guarantee $(u_1, d_1) \leadsto_{\hat{G}_{\scv}} (u_2, d_2)$, which shows the implication in the other direction, and therefore we have
\begin{equation}\label{eq:scq}
(u_1, d_1) \leadsto_{\hat{G}_{\scv}} (u_2, d_2)\iff \scqry(u_1, d_1, u_2, d_2).
\end{equation}
We can now answer a same-context query by answering a reachability query on $\hat{G}_{\scv}$, which can be done by a simple depth-first search (DFS). Let $m$ be the size of the function $fg(u_1)$, then the runtime cost for the query is bounded by $O(m\cdot D^2)$.
\item In $\hat{G}_{\valid}$, if we consider a path $P$ from $(u_1, d_1)$ to $(u_2, d_2)$, then its edges are either intraprocedural edges, summary edges, or call-start edges. As shown above, path segments consisting solely of the first two types correspond to SCVPs, whereas the call-start edges correspond to changing control from a caller function $f_i$ to a called function $f_j$. However, since all exit-return-site edges are removed from $\hat{G}_{\valid}$, the path never returns from $f_j$, but it may continue to call other functions. We say that such a call to $f_j$ is \emph{persistent}. This notion is further formalized in the next chapter. We see that the call and return sequence corresponding to $P$ is consistent with the grammar characterizing IVPs in Chapter \ref{sec:ifds}, and therefore we get that $$(u_1, d_1) \leadsto_{\hat{G}_{\valid}} (u_2\emph{v}, d_2)\implies \qry(u_1, d_1, u_2, d_2).$$

To see the other direction, we note that each IVP can be broken into SCVPs separated by persistent calls. This is expressed more formally in Equation~\eqref{eq:can} in the next Chapter. The same-context paths are enabled in $\hat{G}_{\valid}$ through intraprocedural and summary edges, and persistent calls are enabled through call-start edges, and therefore we obtain
\begin{equation}\label{eq:qry}
(u_1, d_1) \leadsto_{\hat{G}_{\valid}} (u_2, d_2)\iff \qry(u_1, d_1, u_2, d_2).
\end{equation}
This implies that we can answer general interprocedural queries in $O(n\cdot D^2)$ time by a reduction to simple reachability.
\end{itemize}

The discussion above already gives us a solution to our problem: We first compute summaries in the preprocessing phase to obtain $\hat{G}$, construct $\hat{G}_{\valid}$ from it, and answer queries by checking reachability in $\hat{G}_{\valid}$. This has a preprocessing time of $O(n\cdot D^3)$ and a query time of $O(n\cdot D^2)$. Of course, this query time does not scale to a large number of queries and a large number of lines. Our goal is to ultimately reduce the query time without compromising too much on the preprocessing time. We will achieve this in two steps:
\begin{enumerate}[(i)]
\item Utilize the bounded treewidth of control-flow graphs to achieve fast reachability queries on $\hat{G}_{\scv}$, which, by Equation \eqref{eq:scq}, enables efficient querying for \emph{same-context} queries. This is covered in the rest of this chapter.
\item Exploit the previous step along with the bounded treedepth of call graphs to obtain a faster query time for reachability in $\hat{G}_{\valid}$, which, by Equation \eqref{eq:qry}, gives answers to general IFDS queries. This part is presented in Chapter \ref{sec:algo} and is our main technical contribution.
\end{enumerate}

\paragraph{Treewidth of $\hat{G}_{\scv}$~\cite{esop}} For each function $f_i$, define $\hat{G}_{i, \scv}$ to be subgraph of $\hat{G}_{\scv}$ that contains only nodes that lie in $f_i$. From the discussion above, a same-context query $(u_1, d_1, u_2, d_2)$ is answered by a reachability query on $\hat{G}_{fg(u), \scv}$ that checks whether $(u_1, d_1)\leadsto_{\hat{G}_{fg(u), \scv}} (u_2, d_2)$. We will use the provided tree decompositions to answer such reachability queries faster. For this, each function will be processed in isolation. Fix a function $f_i$, and consider its tree decomposition $T_i = (\bags_i, E_{T_i})$. Our algorithm performs the following steps in the preprocessing phase:
\begin{enumerate}[$\text{Phase}$ 1.]
\item \emph{Same-bag reachability}: Precompute all reachability information $(u_1, d_1)\leadsto_{\hat{G}_{i, \scv}} (u_2, d_2)$ for all $u_1, u_2 \in V, d_1, d_2 \in D^*$ where $u_1$ and $u_2$ \emph{simultaneously appear in the same bag in $\bags_i$}, i.e.~there is a $b\in \bags_i$ with $u_1, u_2 \in V_b$.
\item \emph{Ancestor-bag reachability}: Use the answers of the previous step to precompute all reachability information $(u_1, d_1)\leadsto_{\hat{G}_{i, \scv}} (u_2, d_2)$ for all $u_1, u_2 \in V, d_1, d_2 \in D^*$ where there are two bags $b_{1}, b_{2}\in \bags_i$ such that (i)~$u_1\in V_{b_{1}}$, (ii)~$u_2\in V_{b_{2}}$ and (iii)~either $b_1$ is an ancestor of $b_2$ or $b_2$ is an ancestor of $b_1$ in $T_i$.
\end{enumerate}
To answer a same-context query $(u_1, d_1, u_2, d_2)$, we look up a small portion of the precomputed ancestor-bag information for $fg(u_1)$, and conclude from that the answer to the query. We now describe each of these steps in more detail, following the approach of~\cite{esop}.

\paragraph{Phase 1. Same-bag Reachability~\cite{esop}} Consider a bag $b\in \bags_i$. For every $u_1, u_2 \in V_{b}, d_1, d_2 \in D^*$, our goal is to decide whether $(u_1, d_1)\leadsto_{\hat{G}_{i, \scv}} (u_2, d_2)$. If so, we will mark that information by adding a direct edge $((u_1, d_1), (u_2, d_2))$ in $\hat{G}_{i, \scv}$. Denote the edge set of $\hat{G}_{i, \scv}$ by $\hat{E}_{i, \scv}$. The pseudocode in Algorithm~\ref{alg:samebag} describes how to compute all same-bag reachability information. Note that this algorithm will be run $k$ times on $\hat{G}_{i, \scv}$ for all $i\in \{1, \dots, k\}$, one run for every function $f_i\in F$. The algorithm operates on a tree decomposition $T' = (\bags', E_{T'}),$ which is set to $T_i$ when first invoking the algorithm, and does the following:
\begin{compactenum}
\item Pick a leaf bag $b_l\in \bags'$.
\item Run a standard all-pairs graph reachability algorithm on $\hat{G}_{i, \scv}[V_{b_l}\times D^*]$ and for every $(u_1, d_1)\leadsto_{\hat{G}_{i, \scv}[V_{b_l}\times D^*]} (u_2, d_2)$, add the edge $((u_1, d_1), (u_2, d_2))$ to $\hat{E}_{i, \scv}$. In Algorithm~\ref{alg:samebag}, we chose a simple variant of the standard Floyd–Warshall algorithm~\cite{DBLP:journals/cacm/Floyd62a}. We assume the order of iteration over $V_b\times D^*$ is the same in all the loops in lines 9-11.
\item If $b_l$ is not the root of $T'$, then
\item[3.1] recursively solve the problem on $T' - b_l,$ and
\item[3.2] repeat Step 2.
\end{compactenum}

\begin{algorithm}
\caption{Computing same-bag reachability of Phase 1.}\label{alg:samebag}
    \DontPrintSemicolon
    \SetKwFunction{LOC}{SameBagReachability}
    \SetKwFunction{UPD}{UpdataBag}
	Call \LOC{$T_i$}.\;
	\SetKwProg{Fn}{Function}{:}{end}
	 \Fn{\LOC{$T'$}}{
	     Let $b_l$ be a leaf bag in $T'$.\;
	     \UPD{$b_l$}.\;
	     \If{$b_l$ has a parent bag $b_p$}{
	         \LOC{$T'-b_l$}.\;
	         \UPD{$b_l$}.\;
	     }
	 }
	  \Fn{\UPD{$b$}} {
	      \ForEach{$(u_3, d_3)\in V_b\times D^*$}{
	          \ForEach{$(u_1, d_1)\in V_b\times D^*$}{
	              \ForEach{$(u_2, d_2)\in V_b\times D^*$}{
	                  \If{$((u_1, d_1), (u_3, d_3))\in \hat{E}_{i, \scv}$ and $((u_3, d_3), (u_2, d_2))\in \hat{E}_{i, \scv}$}{
	                  	$\hat{E}_{i, \scv}\gets \hat{E}_{i, \scv}\cup ((u_1, d_1), (u_2, d_2))$.\;
	                  }
	              }
	          }
	      }
	  }
\end{algorithm}

\paragraph{Correctness} We show the correctness of the algorithm above by induction on the number of bags in $T'$. If we have one bag, then all paths go through $V_{b_{l}}$ and therefore Step~1 correctly finds all reachability information. Otherwise, suppose our tree decomposition has at least two bags. Define $V_{b_{-l}} = \bigcup_{b'\in \bags', b\neq b_l} V_{b'}$. Consider a path $P$ from $(u_1, d_1)$ to $(u_2, d_2)$ in $\hat{G}_{i, \scv}$ where $u_1, u_2 \in V_{b}$ for some bag $b\in \bags'$. Let $Q$ be the path obtained from $P$ by extracting only the first component in its vertices, i.e.~$Q$'s nodes are in $V_i$. We have 4 cases:
\begin{compactenum}[(i)]
\item $b = b_l$ and $Q$ only traverses nodes in $V_{b_l}$.
\item $b = b_l$ and $Q$ traverses nodes in $V_{b_{-l}}\backslash V_{b_{l}}$.
\item $b \neq b_l$ and $Q$ only traverses nodes in $V_{b_{-l}}$.
\item $b \neq b_l$ and $Q$ traverses nodes in $V_{b_{l}}\backslash V_{b_{-l}}$.
\end{compactenum}
We want to show that in every case, our algorithm adds an edge $((u_1, d_1), (u_2, d_2))$ to $\hat{G}_{i, \scv}$. This holds for case (i) by Step 1, and for case (iii) by the induction hypothesis. Cases (ii) and (iv) are more subtle, since in these cases $P$ spans both $\hat{G}_{i, \scv}[V_{b_l}\times D^*]$ and $\hat{G}_{i, \scv}[V_{b_{-l}}\times D^*]$ of the subproblem solved in Step 3. However, by the cut property, $P$ can only move between those two graphs when $Q$ intersects the set $V_{b_l} \cap V_{b_p}$, where $b_p$ is the parent bag of $b_l$ in $T'$. Because we add new reachability information of $\hat{G}_{i, \scv}[V_{b_l}\times D^*]$ as direct edges in Step 2, then in case (iv), the parts of $Q$ that intersect $V_{b_l}\backslash V_{b_{-l}}$ appear as direct edges in $V_{b_l}\cap V_{b_p}\subseteq V_{-l}$ and therefore there exists another path $P'$ with a $Q'$ that lies completely in $V_{b_{-l}}$. Therefore, by the induction hypothesis, any reachability in $P'$ is recorded by our algorithm. Case (ii) is symmetric to case (iv).

\begin{figure}[H]
	\centering
	\includegraphics[keepaspectratio, scale=0.75]{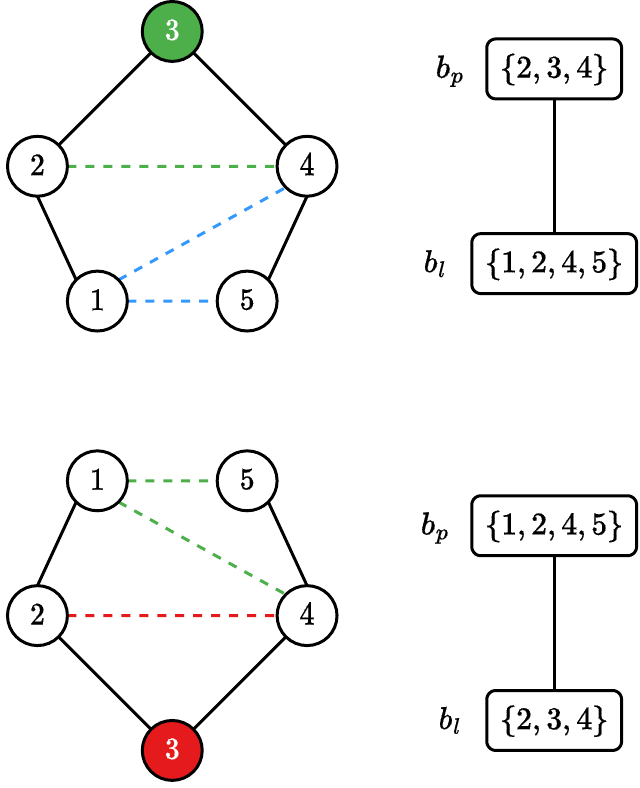}
	\caption{Two graphs showing cases (ii) and (iv) in the correctness proof of Algorithm~\ref{alg:samebag}.}
	\label{fig:samebag}
\end{figure}

\paragraph{Example} We illustrate cases (ii) and (iv) of the correctness proof of Algorithm~\ref{alg:samebag} by two simple graphs in Figure~\ref{fig:samebag}.

\paragraph{Example: Case (ii)} Consider the graph $G_1$ at the top left and its tree decomposition $T_1$ to its right, and suppose we are processing $b_l$. Nodes 1 and 5 are in the same bag $b_l$, and 5 is reachable from 1 in $G_1$. Thus, we want to compute the reachability $1\leadsto_{G_1}5$. A local all-pair reachability computation on $G_1[V_{b_l}]$ will not discover such information because $G_1[V_{b_1}]$ does not contain node 3 in green and hence 1 and 5 are disconnected in $G_1[V_{b_1}]$. However, by the cut property, the path from 1 to 5 can only leave $V_{b_l}$ through nodes in $V_{b_l}\cap V_{b_{p}}$, namely through the nodes 2 and 4. In step 3, the recursive call to $T_1-b_l$ will itself run a local reachability algorithm on $G[V_{b_p}]$ in step 2 that discovers the path $2\leadsto_{G_1}4$, and the dashed green edge $(2, 4)$ will be added as a result. $(2, 4)$ is visible in $G_1[V_{b_l}]$, and therefore after returning from the recursive call to $T_1-b_l$, running a local reachability computation again in step 3.2 will make use of the edge $(2, 4)$ and will find the dashed blue edges that certify $1\leadsto_{G_1}4$ and $1\leadsto_{G_1} 5$.

\paragraph{Example: Case (iv)} Consider $G_2$ at the bottom left of Figure~\ref{fig:samebag} and its tree decomposition $T_2$ to its right, and again suppose we are processing $b_l$. Similar to before, we have $1, 5 \in V_{b_p}$ and hence we want to conclude $1\leadsto_{G_2} 5$, which is not achieved if we run local reachability algorithm on $G_2[V_{b_p}]$ since the node $3\notin V_{b_p}$. However, the local reachability computation in step 2 while processing $b_l$ will discover the dashed red edge $(2, 4)$ which is seen when solving the problem recursively on $b_p$, which will enable step 2 in the subproblem $T_2-b_l$ to discover the dashed green edges $(1, 5)$ and $(1, 4)$ as desired.

\paragraph{Runtime} The algorithm above performs one traversal on the tree decomposition, and in each bag runs two all-pairs reachability computations on a graph of $O(k_1\cdot D)$ nodes, which can be done in $O(k_1^3\cdot D^3)$ time. We treat $k_1$ as a constant and therefore the final runtime to run this algorithm for all $f_i \in F$ is $O(n\cdot D^3)$.

\paragraph{Phase 2. Ancestor-bag Reachability~\cite{esop}} For a bag $b\in \bags_i$, define $anc(b)\subseteq \bags_i$ to be the set of ancestor bags of $b$ in $T_i$, excluding $b$ itself. For every pair of bags $b\in \bags_i, b'\in anc(b)$, we aim to compute $(u_1, d_1)\leadsto_{\hat{G}_{i, \scv}} (u_2, d_2)$ and $(u_2, d_2)\leadsto_{\hat{G}_{i, \scv}}(u_1, d_1)$ for all $u_1 \in V_b, u_2 \in V_{b'}, d_1, d_2 \in D^*$. Similar to same-bag reachability, we record such information as direct edges in $\hat{G}_{i, \scv}$. This is described in Algorithm~\ref{alg:ancbag}. Again, the algorithm will be run $k$ times for every function.

We traverse the tree decomposition top-down and we skip processing the root because all of its ancestor-bag reachability information is already calculated by the previous phase (lines 1-3). Suppose we are processing bag $b$ with parent bag $b_p$, then for every $u_1\in V_b, u_2 \in V_b \cap V_{b_p}, b'\in anc(b), u_3\in V_{b'}, d_1, d_2, d_3\in D^*$, if the edges $((u_1, d_1), (u_2, d_2))$ and $((u_2, d_2), (u_3, d_3))$ are both in $\hat{G}_{i, \scv}$, we add $((u_1, d_1), (u_3, d_3))$ to $\hat{E}_{i, \scv}$ (lines 9-10). Further, we add the edge $((u_3, d_3), (u_1, d_1))$ if the edges $((u_3, d_3), (u_2, d_2))$ and $((u_2, d_2), (u_1, d_1))$ are both present $\hat{G}_{i, \scv}$ (lines 11-12).

\begin{algorithm}
\caption{Computing ancestor-bag reachability of Phase 2.}\label{alg:ancbag}
    \DontPrintSemicolon
    \ForEach{$b\in \bags_i$ in top-down order}{
    	\If{$b$ is the root of $T_i$} {continue.\;}
        Let $b_p$ be the parent of $b$.\;
        \ForEach{$(u_1, d_1)\in V_b\times D^*$}{
            \ForEach{$(u_2, d_2)\in (V_b\cap V_{b_p})\times D^*$}{
            	\ForEach{$b'\in anc(b)$}{
            		\ForEach{$(u_3, d_3)\in V_{b'}$}{
            			\If{$((u_1, d_1), (u_2, d_2))\in \hat{E}_{i, \scv}$ and $((u_2, d_2), (u_3, d_3))\in \hat{E}_{i, \scv}$}{
            				$\hat{E}_{i, \scv}\gets \hat{E}_{i, \scv}\cup((u_1, d_1), (u_3, d_3))$.\;
            			}
		       			\If{$((u_3, d_3), (u_2, d_2))\in \hat{E}_{i, \scv}$ and $((u_2, d_2), (u_1, d_1))\in \hat{E}_{i, \scv}$}{
		       				$\hat{E}_{i, \scv}\gets \hat{E}_{i, \scv}\cup((u_3, d_3), (u_1, d_1))$.\;
		       			}
            		}
                }
            }
        }
    }
\end{algorithm}

\begin{figure}[H]
	\centering
	\includegraphics[keepaspectratio, scale=0.75]{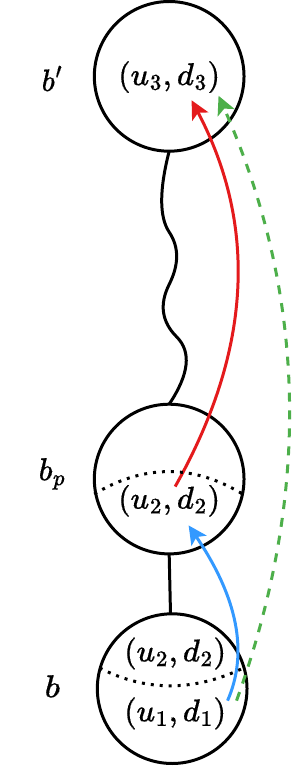}
	\caption{An illustration of computing ancestor-bag reachability}
	\label{fig:ancbag}
\end{figure}

\paragraph{Correctness} The algorithm above correctly computes all the ancestor-bag reachability information of interest. We show this by induction on the number of bags processed so far. At bag $b$, consider a path $P$ from $(u_1, d_1)$ to  $(u_3, d_3)$. Paths from $(u_3, d_3)$ to $(u_1, d_1)$ are analyzed similarly. If $u_1\in V_b\cap V_{b_p}$, then by the induction hypothesis, $(u_1, d_1)\leadsto_{\hat{G}_{i, \scv}} (u_3, d_3)$ has been computed in previous iterations of the tree decomposition traversal. Otherwise $u_1\in V_b\backslash V_{b_p}$. By the cut property, $P$ can only leave $V_b\times D^*$ and reach $(u_3, d_3)$ through a node $(u_2, d_2)$ where $u_2\in V_b\cap V_{b_p}$. See Figure~\ref{fig:ancbag} for a better illustration. The vertices $u_1$ and $u_2$ are in the same bag $V_b$, and hence the reachability $(u_1, d_1)\leadsto_{\hat{G}_{i, \scv}} (u_2, d_2)$ has been marked by the same-bag reachability algorithm as an edge $((u_1, d_1), (u_2, d_2))$ in $\hat{G}_{i, \scv}$ (blue). Moreover, since $u_2$ lies in $V_b\cap V_{b_p}$, then by the induction hypothesis, the reachability $(u_2, d_2)\leadsto_{\hat{G}_{i, \scv}} (u_3, d_3)$ must have been recorded at a previous iteration as an edge $((u_2, d_2), (u_3, d_3))$ (red). Hence, our algorithm will correctly add the edge $((u_1, d_1), (u_3, d_3))$ (green) to record $(u_1, d_1)\leadsto_{\hat{G}_{i, \scv}} (u_3, d_3)$.

\paragraph{Runtime} The algorithm above traverses every bag once and at every bag $b$, it performs $O(k_1^3\cdot D^3 \cdot |anc(b)|)$ work. Since the tree decomposition is balanced, we have $|anc(b)| = O(\lg n)$. Again, treating $k_1$ as a constant, we get a total runtime of $O(n\cdot D^3\cdot \lg n)$ for processing all $f_i \in F$. Using word tricks that exploit the RAM model with word size $\Theta(\lg n)$, we can represent reachability information as a string of bits, and make use of constant-time bit operations to do manipulations that otherwise took $O(\lg n)$ time. This enables us to eliminate the $\lg n$ factor in the runtime and obtain a runtime of $O(n\cdot D^3)$. See~\cite{esop} for details of bit tricks.

\begin{figure}[H]
	\centering
	\includegraphics[keepaspectratio, scale=0.75]{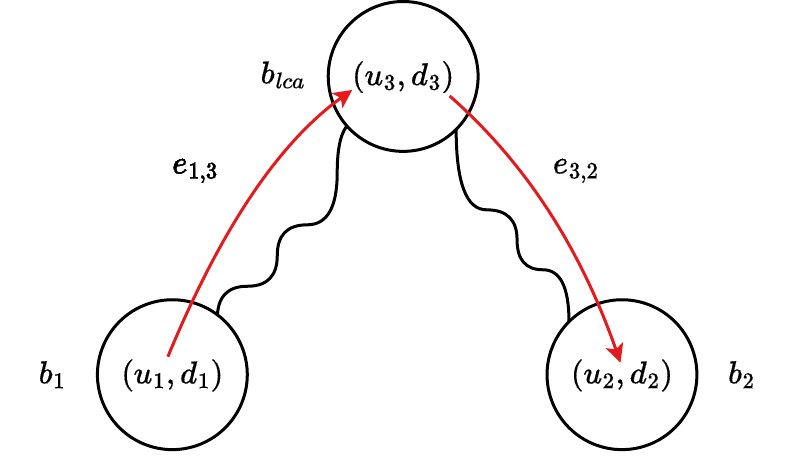}
	\caption{An illustration of answering a same-context query}
	\label{fig:scq}
\end{figure}

\paragraph{Same-context Query~\cite{esop}} Finally, we are ready to present our approach for answering a same-context query using the information saved in our preprocessing, which is shown in Algorithm~\ref{alg:scq}. Suppose we are given a same-context query $(u_1, d_1, u_2, d_2)$ and aim to decide whether $\scqry(u_1, d_1, u_2, d_2).$ If $fg(u_1)\neq fg(u_2)$, we return false, since it is impossible to have a same-context path that starts in a function and ends in a different function. Otherwise, suppose $fg(u_1) = fg(u_2) = f_i$. Recall that by Equation~\eqref{eq:scq}, our task is now reduced to checking if $(u_1, d_1)\leadsto_{\hat{G}_{i, \scv}} (u_2, d_2).$ We first find arbitrary bags $b_1, b_2\in \bags_{i}$ such that $u_1\in V_{b_1}$, $u_2\in V_{b_2}$. We compute the least common ancestor of $b_1$ and $b_2$ in $T_i$ and denote it by $b_{lca}$. We iterate over all $(u_3, d_3)\in V_{b_{lca}}\times D^*$ and check if the edges $e_{1, 3} := ((u_1, d_1), (u_3, d_3))$ and $e_{3, 2} := ((u_3, d_3), (u_2, d_2))$ are both present in $\hat{E}_{i, \scv}$. We return true if and only if the check passes for some $(u_3, d_3)$. See Figure~\ref{fig:scq}.

\paragraph{Correctness} To see why Algorithm~\ref{alg:scq} correctly decides $\scqry(u_1, d_1, u_2, d_2)$, apply the cut property on the edge $(b', b_{lca})$, where $b'$ is on the path from $b_1$ to $b_2$ in $T_i$. We get that a path from $(u_1, d_1)$ to $(u_2, d_2)$ must pass through a node $(u_3, d_3)$ for $u_3\in V_{b_{lca}}$. Such path can be broken into two paths: the first is $P_{1, 3}$ from $(u_1, d_1)$ to $(u_3, d_3)$, and the second is $P_{3, 2}$ from $(u_3, d_3)$ to $(u_2, d_2)$. Therefore, it suffices to check all such $(u_3, d_3)$ and see if such $P_{1, 3}$ and $P_{3, 2}$ exist. Note that by definition, $b_{lca}$ is an ancestor of both $b_1, b_2$ and hence if paths $P_{1, 3}, P_{3, 2}$ do exist, then $\hat{G}_{i, \scv}$ is guaranteed to contain the corresponding $e_{1, 3}$ and $e_{3, 2}$ edges because of our ancestor-bag preprocessing .
\begin{algorithm}
\caption{Answering a same-context query.}\label{alg:scq}
    \DontPrintSemicolon
    \SetKwFunction{SCQ}{SameConextQuery}
    \SetKwFunction{UPD}{UpdataBag}
	\SetKwProg{Fn}{Function}{:}{end}
	 \Fn{\SCQ{$u_1, d_1, u_2, d_2$}}{
	 	\If{$fg(u_1)\neq fg(u_2)$} { return false.\;}
	 	Suppose $fg(u_1) = f_i$.\;
	 	Pick an arbitrary bag $b_1\in\bags_{i}$ s.t. $u_1\in V_{b_1}$.\;
	 	Pick an arbitrary bag $b_2\in\bags_{i}$ s.t. $u_2\in V_{b_2}$.\;
	 	$b_{lca}\gets$ the least common ancestor of $b_1$ and $b_2$ in $T_i$.\;
	 	\ForEach{$(u_3, d_3)\in V_{b_{lca}}\times D^*$}{
	 		\If{$((u_1, d_1), (u_3, d_3))\in \hat{E}_{i, \scv}$ and $((u_3, d_3), (u_2, d_2))\in \hat{E}_{i, \scv}$}{
	 			return true.\;
	 		}
	 	}
	 	return false.\;
	 }
\end{algorithm}

\paragraph{Runtime} We remark that the least common ancestor queries can be answered in $O(1)$ time with $O(n)$ cost in the preprocessing phase \cite{DBLP:journals/siamcomp/HarelT84}. Therefore, the query time is $O(k_1\cdot D) = O(D)$. Again, we treat $k_1$ as a constant and use word tricks to represent reachability information more succinctly and achieve a runtime of $O(\lceil \frac{D}{\lg n}\rceil)$~\cite{esop}.

\chapter{Parameterized Algorithms for IFDS} \label{sec:algo}

In this chapter, we present our parameterized algorithm for solving the general case of IFDS data-flow analysis, assuming that the control-flow graphs have bounded treewidth and the call graph has bounded treedepth.

\paragraph{Algorithm for Same-Context IFDS} As discussed in the previous chapter, the work~\cite{esop} provides an on-demand parameterized algorithm for same-context IFDS. This algorithm requires a balanced and binary tree decomposition of constant width for every control-flow graph and provides a preprocessing runtime of $O(n \cdot D^3),$ after which it can answer \emph{same-context} queries in time $O \left( \lceil \frac{D}{\lg n} \rceil \right).$ Recall that a same-context query $(u_1, d_1, u_2, d_2)$ is only concerned with SCVPs from $(u_1, d_1)$ to $(u_2, d_2)$ which form a restricted subset of the IVPs we are interested in. Nonetheless, we use~\cite{esop}'s algorithm for same-context queries as a black box. 

\paragraph{Stack States} A \emph{stack state} is simply a finite sequence of functions $\xi = \langle \xi_i \rangle_{i=1}^s \in F^s.$ Recall that $F$ is the set of functions in our program. We use a stack state to keep track of the set of functions that have been called but have not finished their execution and returned yet.

\paragraph{Persistence and Canonical Partitions} Consider an IVP $\Pi = \langle \pi_i \rangle_{i=1}^p$ in the supergraph $G$ and let $\Pi^* = \langle \pi^*_i \rangle_{i=1}^s$ be the sub-sequence of $\Pi$ that only includes call vertices $c_l$ and return vertices $r_l.$ For each $\pi^*_i$ that is a call vertex, let $f_{x_i}$ be the function called by $\pi_i^*.$ We say the function call to $f_{x_i}$ is \emph{temporary} if $\pi^*_i$ is matched by a corresponding return-site vertex $\pi^*_j$ in $\Pi^*$ with $j > i.$ Otherwise, $f_{x_i}$ is is a \emph{persistent} function call. In other words, temporary function calls are the ones that return before the end of the path $\Pi$ and persistent ones are those that are added to the stack but never popped. So, if the stack is at state $\xi$ before executing $\Pi,$ it will be in state $\xi \cdot \langle f_{x_{i_1}} \cdot f_{x_{i_2}} \cdots f_{x_{i_r}} \rangle$ after $\Pi$'s execution, in which $f_{x_{i_1}}, \dots, f_{x_{i_r}}$ are our persistent function calls. Moreover, we can break down the path $\Pi$ as follows:
\begin{equation} \label{eq:can}
\Pi = \Sigma_0 \cdot \Sigma_1 \cdot \pi_{i_1} \cdot \Sigma_2 \cdot \pi_{i_2}  \cdots \Sigma_{r} \cdot \pi_{i_r} \cdot \Sigma_{r+1} 
\end{equation}
in which $\Sigma_0$ is an \emph{intraprocedural} path, i.e.~a part of $\Pi$ that remains in the same function. Note that we either have $\Pi = \Sigma_0$ or $\Sigma_0$ should end with a function call. For every $j \neq 0,$ $\Sigma_j$ is an SCVP from the starting point of a function and $\pi_{i_j}$ is a call vertex that calls the next persistent function $f_{x_{i_j}}.$ We call~\eqref{eq:can} the \emph{canonical partition} of the path $\Pi.$

\begin{figure}
	\centering
	\includegraphics[scale=.7]{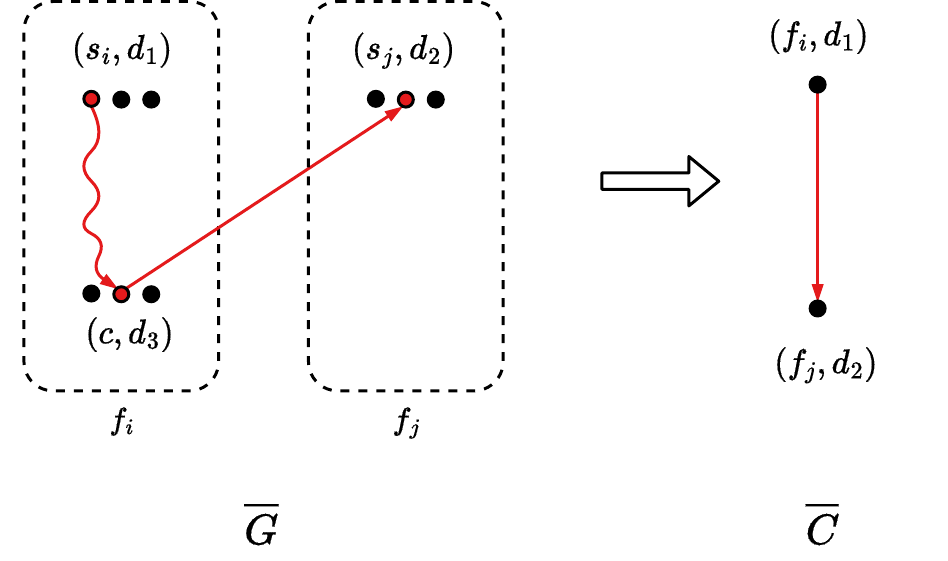}
	\caption{An illustration of the exploded call graph's construction}
	\label{fig:ccc}
\end{figure}

\paragraph{Exploded Call Graph} Let $C = (F, E_C)$ be the call graph of our IFDS instance, in which $F$ is the set of functions in the program. We define the \emph{exploded call graph} $\overline{C} = (\overline{F}, \overline{E_C})$ as follows:
\begin{itemize}
	\item Our vertex set $\overline{F}$ is simply $F \times D^*.$ Recall that $D^* := D \cup \{\zero\}.$
	\item There is an edge from the vertex $(f_i, d_1)$ to the vertex $(f_j, d_2)$ in $\overline{E_C}$ iff:
	\begin{compactitem}
		\item There is a call statement $c \in V$ in the function $f_i$ that calls $f_j$, i.e. $(f_i, f_j)\in E_C$; 
		\item There exist a data fact $d_3 \in D^*$ such that (i)~there is an SCVP from $(s_{i}, d_1)$ to $(c, d_3)$ in the exploded supergraph $\overline{G},$ and (ii)~there is an edge from $(c, d_3)$ to $(s_{j}, d_2)$ in $\overline{G}.$ 
	\end{compactitem}
\end{itemize}
An illustration of how edges are added to the exploded call graph is shown in Figure~\ref{fig:ccc}. The red path segment in $\overline{G}$ on the left results in adding the red edge to $\overline{C}$ on the right. The edges of the exploded call graph model the effect of an IVP that starts at $s_{i},$ i.e.~the first line of $f_i,$ when the function call stack is empty and reaches $s_j,$ with stack state $\langle f_j \rangle$. Informally, this corresponds to executing the program starting from $f_i,$ potentially calling any number of temporary functions, then waiting for all of these temporary functions and their children to return so that we again have an empty stack, and then finally calling $f_j$ from the call-site $c$, hence reaching stack state $\langle f_j \rangle.$  Intuitively, this whole process models the substring $\Sigma \cdot c$ in the canonical partition of a valid path, in which $\Sigma$ is an SCVP, and $f_j$ is the next persistent function, which was called at $c$. Hence, going forward, we do not plan to pop $f_j$ from the stack.

\begin{figure}
	\begin{center}
	\includegraphics[scale=0.75]{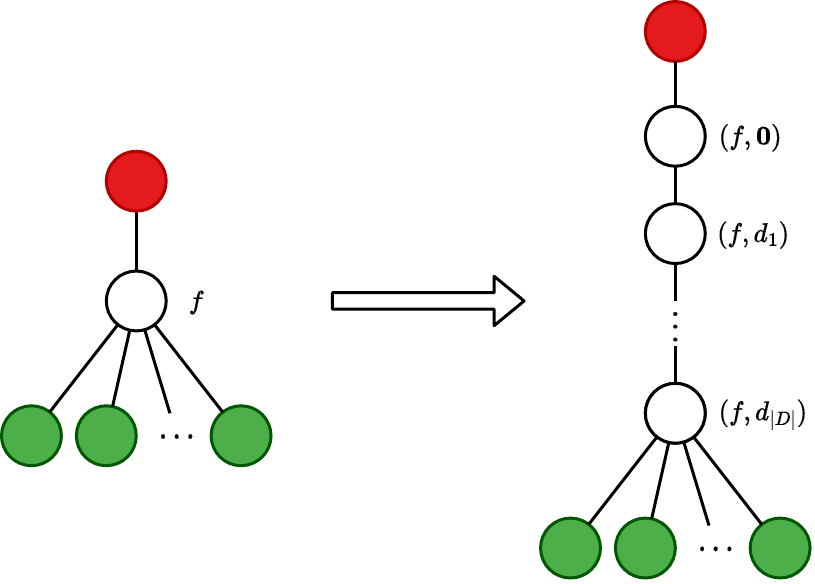}
	\end{center}
	\caption{Obtaining $\overline{T}$ from $T$ by expanding each vertex to a path.}
	\label{fig:path}
\end{figure}

\paragraph{Treedepth of $\overline{C}$} Recall that we have a POT $T = (F, E_T)$ for the call graph $C$ with root $r_{T}$ and depth $k_2.$  In $\overline{C},$ every  $f \in C$ is replaced by $|D^*|$ vertices $(f, \zero), (f, d_1), \ldots, (f, d_{|D|}).$ We can obtain a valid POT $\overline{T} = (\overline{F}, \overline{E_{T}})$ with root $(r_T, \zero)$ for $\overline{C}$ by processing the POT $T$ in a top-down order and replacing every vertex that corresponds to a function $f$ with a path of length $|D^*|,$ as shown in Figure~\ref{fig:path}. It is straightforward to verify that $\overline{T}$ is a valid POT of depth $k_2 \cdot |D^*|$ for $\overline{C}.$

\begin{figure}[H]
	\begin{center}
		\includegraphics[scale=0.75]{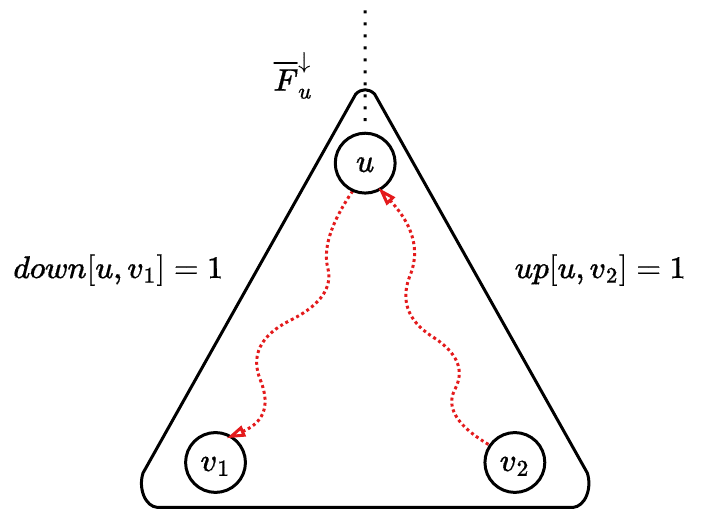}
	\end{center}
	\caption{An illustration of $up$ and $down.$}
	\label{fig:updownex}
\end{figure}

\paragraph{Reachability on $\overline{C}$ via $\overline{T}$} The query phase of our algorithm relies on efficiently answering standard reachability queries in the exploded call graph $\overline{C}.$ To achieve this, we will exploit the POT $\overline{T}$ for $\overline{C}$. For every vertex $u$ in $\overline{T},$ let $\subtree_u$ be the subtree of $\overline{T}$ rooted at $u$ and $\descendants_u$ be the set of descendants of $u$. Note that here $u$ stands for a pair of the form $(f, d)\in F \times D^*$, and should not be confused as a node of the supergraph $G$. For every $u$ and every $v \in \descendants_u,$ define $up[u, v]$ and $down[u, v]$ as follows:
$$
~~~~up[u, v] := \left\{\begin{matrix}
	1	& ~~~~~ & \text{there is a path from } v \text{ to } u \text{ in } \overline{C}[\descendants_u] \\
	0	& & \text{otherwise} \\
\end{matrix}\right. ;
$$
$$
down[u, v] := \left\{\begin{matrix}
	1	& ~~~~~ & \text{there is a path from } u \text{ to } v \text{ in } \overline{C}[\descendants_u] \\
	0	& & \text{otherwise} \\
\end{matrix}\right. .
$$
Note that in the definition above, we are only considering paths whose internal vertices are in the subtree of $u$. See Figure~\ref{fig:updownex} for further illustration. Here, we have a path from $u$ to $v_1\in \descendants_u$ that is inside $\overline{C}[\descendants_u],$ and therefore we have $down[u, v_1]=1.$ We similarly get $up[u, v_2]=1.$ Now we show that $up$ and $down$ give us sufficient information to check if $u\leadsto_{\overline{C}} v$ for $u, v\in \overline{F}$.

\begin{figure}
	\begin{center}
	\includegraphics[scale=0.75]{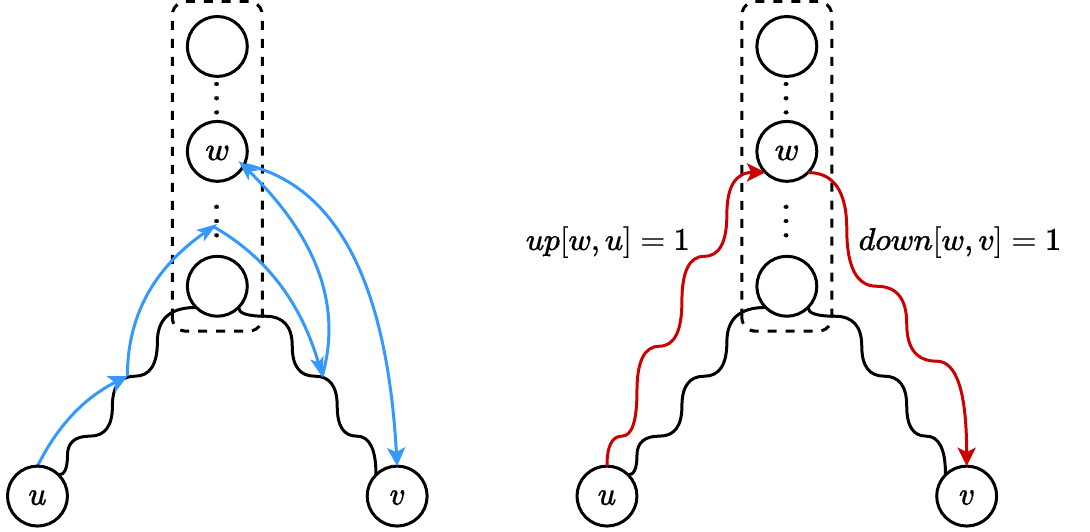}
	\end{center}
	\caption{The correspondence between $P$ and the values of $up$ and $down$.}
	\label{fig:updown}
\end{figure}

Suppose $P$ is a path in our exploded call graph $\overline{C}$ from vertex $u$ to a vertex $v$, and consider its trace in the POT $\overline{T}$. See the left-hand side of Figure~\ref{fig:updown}, where the blue edges indicate edges of $P$. Let $A\subseteq \overline{F}$ be the set of common ancestors of $u$ and $v$ in $\overline{T}$, denoted by the dashed rectangle. By the path property of POTs, we know that $P$ has to go through some ancestor node in $A$. Further, one of these ancestors that lie on $P$ has the smallest depth, let it be $w$. The path $P$ can be broken into two concatenated paths: $P_1$ from $u$ to $w$ and $P_2$ from $w$ to $v$. We claim that all internal vertices of both $P_1$ and $P_2$ lie entirely in $\overline{C}[\descendants_w]$. To see why this is the case, note that by definition of a POT, these paths can only leave $\overline{C}[\descendants_w]$ by going through an ancestor $w'\in A$ that is outside $\overline{C}[\descendants_w].$ However, since $w$ is the ancestor of smallest depth, such $w'$ does not exist and thus our claim is established. By definition of $up$ and $down$, we must have: \begin{equation}
\label{eq:updown} up[w, u] = 1\land down[w, v] = 1,
\end{equation}
which is illustrated in Figure~\ref{fig:updown} to the right. Moreover, if no $w\in A$ satisfies Equation~\eqref{eq:updown}, then there is no path $P$ from $u$ to $v$ in $\overline{C}$. Hence, $P$ exists iff there is some ancestor $w$ satisfying Equation~\eqref{eq:updown} and we get the following correspondence: 
\begin{equation}\label{eq:expcallqry}
u\leadsto_{\overline{C}} v\iff \bigvee_{w\in A} \left(up[w, u] = 1\land down[w, v] = 1\right).
\end{equation}

We are now ready to present our algorithm, which consists of a multi-phase preprocessing step followed by a query step in which it can efficiently answer general interprocedural queries.

\paragraph{Preprocessing} The preprocessing phase of our algorithm consists of the four steps described below. Pseudocode for the first three steps is given in Algorithm~\ref{alg:steps1234} whereas the fourth step is shown in Algorithm~\ref{alg:ancreach}.
\begin{enumerate}[(Step 1)]
	\item \textbf{\emph{Same-context Preprocessing:}} Our algorithm runs the preprocessing algorithm of Chapter~\ref{sec:review} for same-context IFDS. This is done as a black box. 
	\item \textbf{\emph{Intraprocedural Preprocessing:}} For every exploded supergraph vertex $(u, d) \in \overline{V},$ for which $u$ is a line of the program in the function $f_i\in F$, our algorithm performs an \emph{intraprocedural} reachability analysis and finds a list $I_{u, d}$ of all the vertices of the form $(c, d')$ such that:
	\begin{compactitem}
		\item $c$ is a call-site vertex in the same function $f_i.$
		\item There is an \emph{intraprocedural} path from $(u, d)$ to $(c, d')$ that always remains within $f_i$ and does not cause any function calls.
	\end{compactitem}
	Our algorithm computes this by a simple reverse DFS on $\overline{G}[V_i \times D^*]$ from every $(c, d').$ Recall that $V_i$ denotes the supergraph nodes in the control-flow graph of $f_i$, and thus $\overline{G}[V_i \times D^*]$ denotes a restriction of the exploded supergraph to nodes (with first component) in $f_i$. This is done in lines 4-8. Intuitively, this step is done so that we can later handle the first part, i.e.~$\Sigma_0,$ in the canonical partition of Equation~\eqref{eq:can}. Note that this step is entirely intraprocedural and our reverse DFS is equivalent to the classical algorithms of~\cite{kildall1972global}. Moreover, we can consider $\Sigma_0$ to be an SCVP instead of merely an intraprocedural path. In this case, we can rely on same-context queries of Chapter~\ref{sec:review} to do this step of our preprocessing.
	\item \textbf{\emph{Computing the Exploded Call Graph:}} Our algorithm generates the exploded call graph $\overline{C}$ using its definition above which was illustrated in Figure~\ref{fig:ccc}. It iterates over every function $f_i$ and call site $c$ in $f_i.$ Let $f_j$ be the function called at $c.$ For every pair $(d_1, d_3) \in D^* \times D^*,$ our algorithm queries the same-context IFDS framework of Chapter~\ref{sec:review} to see if there is an SCVP from $(s_i, d_1)$ to $(c, d_3).$ Note that we can make such queries since we have already performed the required same-context preprocessing in Step 1 above. If the query's result is positive, the algorithm iterates over every $d_2 \in D^*$ such that $((c, d_3), (s_{j}, d_2))$ is an edge in the exploded supergraph $\overline{G},$ and adds an edge from $(f_i, d_1)$ to $(f_j, d_2)$ in $\overline{C}.$ This is done in lines 9-16. The algorithm also computes the POT $\overline{T}$ as mentioned above, which is done in lines 17-23. Intuitively, this step allows us to summarize the effects of each function call in the call graph so that we can later handle the control-flow graphs and the call graph separately.
\end{enumerate}
\begin{algorithm}
\caption{Steps 1-4.}\label{alg:steps1234}
    \DontPrintSemicolon
    \SetKwFunction{SCQ}{SameConextQuery}
    Run the preprocessing algorithms of Chapter~\ref{sec:review}. \tcp{Step 1.}
    \tcc{We now assume access to \SCQ that answers same-context queries in $O(\lceil \frac{D}{\lg n}\rceil)$ time.}
    \tcp{Step 2.}
    \ForEach{$(u, d)\in \overline{V}$} {
    	$I_{u, d}\gets \emptyset$.\;
    }
    
    \ForEach{$f_i \in F$}{
    	\ForEach{$(c, d')\in V_i\times D^*$ s.t. $c$ is a call node}{
    		Run reverse DFS on $\overline{G}[V_i\times D^*]$ from $(c, d')$ to obtain $R_{c, d'} = \{(u, d) \in V_i\times D^* ~|~(u, d)\leadsto_{\overline{G}[V_i\times D^*]}(c, d')\}$.\;
    		\ForEach{$(u, d)\in R_{c, d'}$}{
    			$I_{u, d}\gets I_{u, d} \cup (c, d')$.\;
    		}
    	}
    }
    \tcp{Step 3.}
	$\overline{C}\gets (\overline{F\emph{}}, \emptyset).$ \tcp{Edges of $\overline{C}$ are denoted with $\overline{E_C}$.}
	\ForEach{$f_i\in F$}{
    	\ForEach{$c\in V_i$ s.t. $c$ is a call node}{
    	    Suppose $f_j$ is the functioned called at $c$.\;
    		\ForEach{$(d_1, d_3)\in D^*\times D^*$}{
    			\If{\SCQ{$s_i, d_1, c, d_3$}$=$ true}{
    				\ForEach{$d_2\in D^*$ s.t. $((c, d_3), (s_{j}, d_2))\in \overline{E}$}{
    					$\overline{E_C}\gets \overline{E_C}\cup ((f_i, d_1), (f_j, d_2))$.\;
    				}
    			}
    		}
    		
    	}
	}
	$\overline{T} \gets (\overline{F}, \emptyset).$ \tcp{Edges of $\overline{T}$ are denoted with $\overline{E_T}$.}
	
	\ForEach{$f_i \in F$}{
		$\overline{E_T}\gets \overline{E_T}\cup ((f_i, \zero), (f_i, d_1))$.\;
		\ForEach{$j\in\{1, \dots, |D|-1\}$}{
			$\overline{E_T}\gets \overline{E_T}\cup ((f_i, d_j), (f_i, d_{j+1}))$.\;
		}
	}
	\ForEach{$(f_i, f_j)\in E_T$}{
		$\overline{E_T}\gets \overline{E_T}\cup ((f_i, d_{|D|}), (f_j, \zero))$.\;
	}
	
\end{algorithm}

\begin{figure}
	\begin{center}
	\includegraphics[scale=0.75]{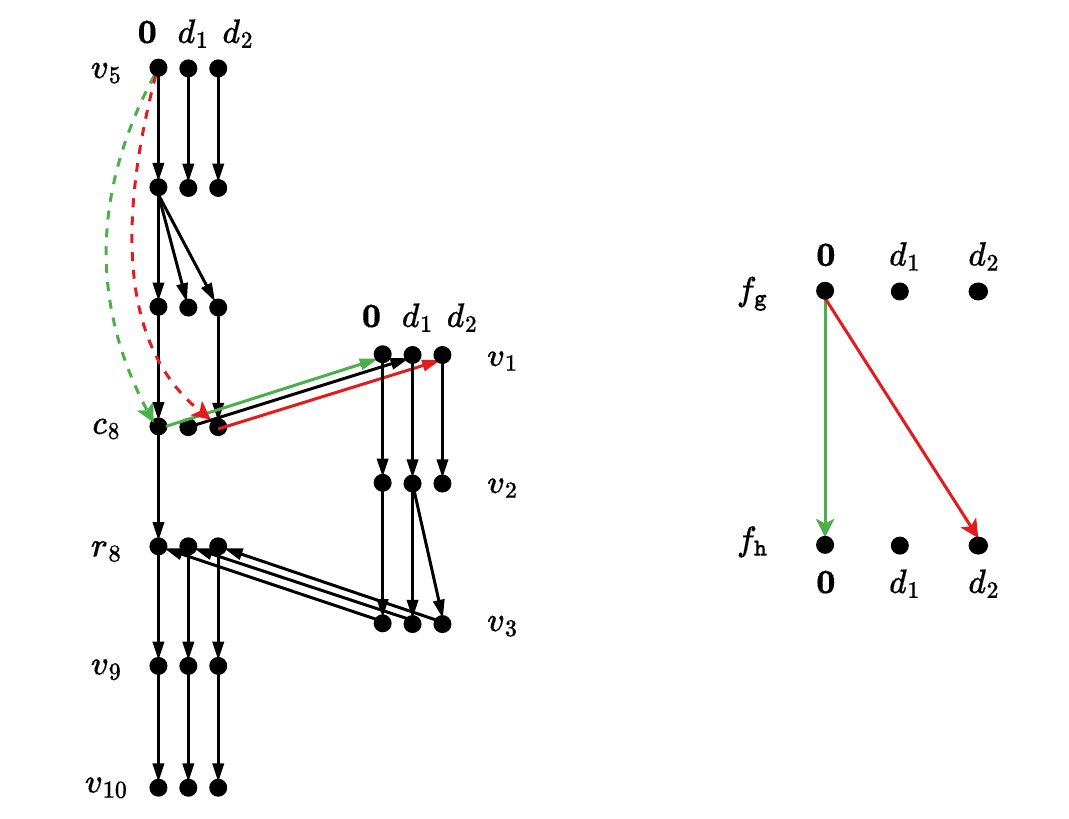}
	\end{center}
	\caption{To the left is an exploded supergraph and to the right is its exploded call graph.}
	\label{fig:step23}
\end{figure}

\paragraph{Example} Consider again the exploded supergraph of Figure~\ref{fig:exploded}. There is only one call node $c_8$, and after running step 2 of our algorithm, we will have:
\begin{align*}	
&I_{v_5, \zero} = I_{v_6, \zero} = \{(c_8, \zero), (c_8, d_2)\},I_{v_7, \zero} = I_{c_8, \zero} = \{(c_8, \zero)\},\\
&I_{c_8, d_1} = \{(c_8, d_1)\}, I_{v_7, d_2} = I_{c_8, d_2} = \{(c_8, d_2)\},\\
&I_{v, d} = \emptyset \text{ for all other nodes } (v, d).
\end{align*}
After running step 3, we will get the exploded call graph on the right-hand side of Figure~\ref{fig:step23}. On the left, there is an SCVP from $(v_5, \zero)$ to $(c_8, \zero)$ denoted by the dashed green edge, and there is an edge from $(c_8, \zero)$ to $(v_1, \zero)$. Both of these facts lead us to add the green edge $(f_{\texttt{g}}, \zero)$ to $(f_{\texttt{h}}, \zero)$ in the exploded call graph to the right. Similarly, we add an edge from $(f_{\texttt{g}}, \zero)$ to $(f_{\texttt{h}}, d_2).$

\begin{enumerate}[(Step 1)]\addtocounter{enumi}{3}
	\item \textbf{\emph{Computing Ancestral Reachability in $\overline{T}$:}}
	In this step, we compute $up[u, v]$ and $down[u, v]$ for all $u\in \overline{F}, v\in \descendants_u$ as defined above. This step is shown in Algorithm~\ref{alg:ancreach}. Our algorithm finds the values of $down[u, v]$ by simply running a DFS from $u$ but ignoring all the edges that leave the subtree $\subtree_u$ (lines 5-7). It also finds the values of $up[u, v]$ by a similar DFS in which the orientation of all edges is reversed (lines 8-10).
	 
\end{enumerate}

\begin{algorithm}
\caption{Computing ancestral reachability (step 4).}\label{alg:ancreach}
    \DontPrintSemicolon
\ForEach{$u\in \overline{F}$}{
	\ForEach{$v\in \subtree_u$}{
		$up[u, v] \gets 0$.\;
		$down[u, v] \gets 0$.\;
	}
	Run DFS on $\overline{C}[\descendants_u]$ from $u$ to obtain $R = \{v \in \overline{F} ~|~u\leadsto_{\overline{C}[\descendants_u]}v\}$.\;
	\ForEach{$v\in R$}{
		$down[u, v] \gets 1$.\;
	}
	Run reverse DFS on $\overline{C}[\descendants_u]$ from $u$ to obtain $R_{rev} = \{v \in \overline{F} ~|~v\leadsto_{\overline{C}[\descendants_u]}u\}$.\;
	\ForEach{$v\in R_{rev}$}{
		$up[u, v] \gets 1$.\;
	}
}

\end{algorithm}

\begin{figure}
	\begin{center}
		\includegraphics[scale=0.75]{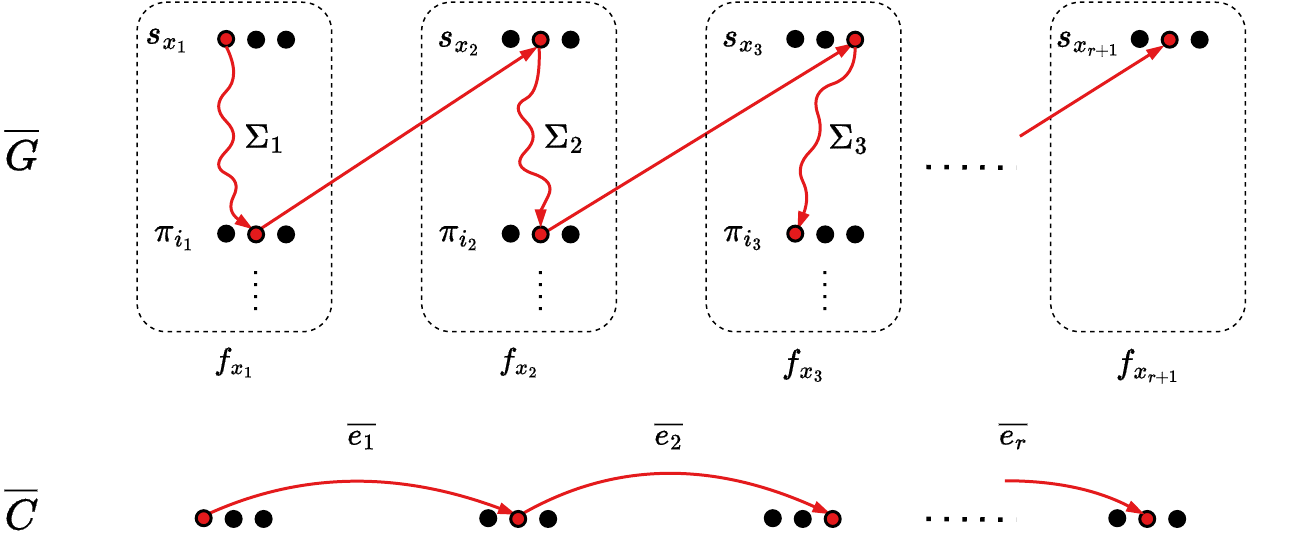}
	\end{center}
	\caption{To the left is an exploded supergraph and to the right is its exploded call graph.}
	\label{fig:nice}
\end{figure}

\paragraph{Query} After the end of the preprocessing phase, our algorithm is ready to accept queries. Suppose that a query $q$ asks whether there exists an IVP from $(u_1, d_1)$ to $(u_2, d_2)$ in $\overline{G}.$ Suppose that $\overline{\Pi}$ is such a valid path and $\Pi$ is its trace on the supergraph $G$, i.e.~the path obtained from $\overline{\Pi}$ by ignoring the second component of every vertex. We consider the canonical partition of $\Pi$ as
$$\Pi = \Sigma_0 \cdot \left( \Sigma_1 \cdot \pi_{i_1} \right) \cdot \left(\Sigma_2 \cdot \pi_{i_2} \right)  \cdots \left( \Sigma_{r} \cdot \pi_{i_r} \right) \cdot \Sigma_{r+1}$$
and its counterpart in $\overline{\Pi}$ as
$$\overline{\Pi} = \overline{\Sigma_0} \cdot \left( \overline{\Sigma_1} \cdot \overline{\pi_{i_1}} \right) \cdot \left( \overline{\Sigma_2} \cdot \overline{\pi_{i_2}} \right) \cdots \left( \overline{\Sigma_{r}} \cdot \overline{\pi_{i_r}} \right) \cdot \overline{\Sigma_{r+1}}.$$ Let $\overline{\Sigma_j}[1]$ be the first vertex in $\overline{\Sigma_j}.$
For every $j \geq 1,$ consider the subpath $$\overline{\Sigma_j} \cdot \overline{\pi_{i_j}} \cdot \overline{\Sigma_{j+1}}[1].$$ 
This subpath starts at the starting point $s_x$ of some function $f_x$ and ends at the starting point $s_{y}$ of the function $f_y$ called in $\overline{\pi_{i_j}}.$ Thus, it goes from a vertex of the form $(s_x, d_1)$ to a vertex of the form $(s_y, d_2).$ However, by the definition of our exploded call graph $\overline{C},$ we must have an edge $\overline{e_j}$ in $\overline{C}$ going from $(f_x, d_1)$ to $(f_y, d_2).$ With a minor abuse of notation, we do not differentiate between $f_x$ and $s_x$ and replace this subpath with $\overline{e_j}$. Hence, every IVP $\overline{\Pi}$ can be partitioned in the following format:
 $$\overline{\Pi} =  \overline{\Sigma_0} \cdot \overline{e_1} \cdot \overline{e_2} \cdots \overline{e_r} \cdot \overline{\Sigma_{r+1}}.$$
In other words, to obtain an IVP, we should first take an intraprocedural path $\overline{\Sigma_0}$ in our initial function, followed by a path $\overline{e_1} \cdot \overline{e_2} \cdots \overline{e_r}$ in the exploded call graph $\overline{C},$ and then an SCVP $\overline{\Sigma_{r+1}}$ in our target function. Note that $\overline{\Sigma_{r+1}}$ begins at the starting point of our target function. Figure~\ref{fig:nice} illustrates this idea: The path segment $$\left( \overline{\Sigma_1} \cdot \overline{\pi_{i_1}} \right) \cdot \left( \overline{\Sigma_2} \cdot \overline{\pi_{i_2}} \right) \cdots \left( \overline{\Sigma_{r}} \cdot \overline{\pi_{i_r}} \right)\cdot \overline{\Sigma_{r+1}}[1],$$ in exploded supergraph (shown in red on the top) has a corresponding path $\overline{e_1} \cdot \overline{e_2} \cdots \overline{e_r}$ in the exploded call graph (shown in red at the bottom).

Our algorithm uses the observation above to answer the queries. Recall that the query $q$ is asking whether there exists a path from $(u_1, d_1)$ to $(u_2, d_2)$ in $\overline{G}$. Let $f_i$ be the function of $u_1$ and $f_j$ be the function containing $u_2$. Our algorithm performs the following steps to answer the query, which are described as pseudocode in Algorithm~\ref{alg:qry}:
\begin{enumerate}
	\item We first check if there is an SCVP from $(u_1, d_1)$ to $(u_2, d_2)$, and if so, we return true (lines 2-3).
	\item Take all vertices of the form $(c, d_3)$ such that $c$ is a call vertex in $f_i$ and $(c, d_3)$ is intraprocedurally reachable from $(u_1, d_1)$ (line 9). This was already precomputed in Step 2 of our preprocessing.
	\item Find all successors of the vertices in Step 2 in $\overline{G}$ (line 10). We only consider successor vertices of the form $(s_{x}, d_4)$ for some function $f_x,$ which have corresponding nodes in the exploded call graph of the form $(f_x, d_4).$ 
	\item Compute the set of all $(f_j, d_5)$ vertices in $\overline{C}$ that are reachable from one of the $(f_x, d_4)$ vertices obtained in the previous step (lines 11-12). In this case, the algorithm uses the path property of POTs and tries all possible common ancestors of $(f_j, d_5)$ and $(f_x, d_4)$ as potential smallest-depth vertices in the path, as discussed above. This is done through a call at line 12 to the helper function $\texttt{ExplodedCallGraphReachability}$ which is a direct implementation of the correspondence in Equation~\eqref{eq:expcallqry}.
	\item For each $(f_j, d_5)$ found in the previous step, ask the same-context query from $(s_{j}, d_5)$ to $(u_2, d_2)$ (line 13). For these same-context queries, our algorithm uses the method of Chapter~\ref{sec:review} as a black box. Since $s_j, u_2, d_2$ are fixed throughout the query, this step is cached for all $d_5\in D^*$ to avoid redundant computation (lines 6-8).
	\item If any of the same-context queries in the previous step return true, then our algorithm also answers true to the query $q$. Otherwise, it answers false. 
\end{enumerate}

\begin{algorithm}
\caption{Answering a query.}\label{alg:qry}
    \DontPrintSemicolon
    \SetKwFunction{SCQ}{SameConextQuery}
    \SetKwFunction{QRY}{Query}
    \SetKwFunction{ECR}{ExplodedCallGraphReachability}
	\SetKwProg{Fn}{Function}{:}{end}
	 \Fn{\QRY{$u_1, d_1, u_2, d_2$}}{
	 	\If{\SCQ{$u_1, d_1, u_2, d_2$}$=true$}{
	 		return true.\;
	 	}
	 	Suppose $fg(u_1) = f_i$.\;
	 	Suppose $fg(u_2) = f_j$.\;
	 	$cache\gets \emptyset$.\;
	 	\ForEach{$d_5\in D^*$}{
	 		$cache \gets cache \cup (s_j, d_5, u_2, d_2,$ \SCQ{$s_j, d_5, u_2, d_2$}$)$
	 	}
	 	\ForEach{$(c, d_3)\in I_{u_1, d_1}$}{
	 		\ForEach{$(s_x, d_4)$ s.t. $((c, d_3),(s_x, d_4))\in \overline{E}$}{
	 			\ForEach{$d_5\in D^*$}{
	 				\If{\ECR{$(f_x, d_4), (f_j, d_5)$} $= true$}{
	 					\If{$(s_j, d_5, u_2, d_2, true)\in cache$}{
	 						return true.\;
	 					}
	 				}
	 			}
	 		}
	 	}
	 	return false.\;
	 }
	 \Fn{\ECR{$u, v$}}{
	 	$A_u\gets$ set of ancestors of $u$ in $\overline{T}$.\;
	 	$A_v\gets$ set of ancestors of $u$ in $\overline{T}$.\;
	 	$A\gets A_u \cap A_v$.\;
	 	\ForEach{$w\in A$}{
	 		\If{$up[w, u] = 1$ and $down[w, v] = 1$}{
	 			return true.\;
	 		}
	 	}
	 	return false.\;
	 }
\end{algorithm}

\begin{figure}
	\begin{center}
		\includegraphics[scale=0.9]{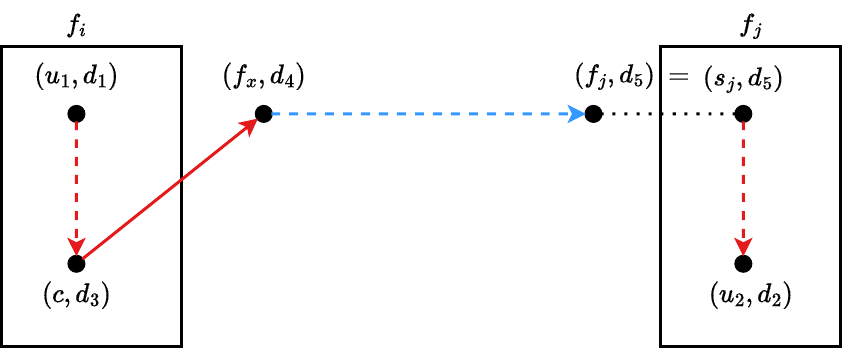}
	\end{center}
\caption{An overview of the query phase.}
\label{fig:ov}
\end{figure}

\paragraph{Intuition} Figure~\ref{fig:ov} provides an overview of how our query phase breaks an IVP down between $\overline{G}$ (red) and $\overline{C}$ (blue). We do not distinguish between the vertex $(f_j, d_5)$ of $\overline{C}$ and vertex $(s_{j}, d_5)$ of $\overline{G}.$ Explicitly, any IVP from $(u_1, d_1)$ to $(u_2, d_2)$ that fails the check at line 2 in Algorithm~\ref{alg:qry} should first begin with an intraprocedural segment in the original function $f_j$. This part is precomputed and shown in red. Then, it switches from the exploded supergraph to the exploded call graph and follows a series of function calls. This is shown in blue. We have already precomputed the effect of each edge in the call graph and encoded this effect in the exploded call graph. Hence, the blue part of the path is simply a reachability query, which we can answer efficiently using our POT through $\texttt{ExplodedCallGraphReachability}$. We would like to see whether there is a path from $a:=(f_x, d_4)$ to $b:=(f_j, d_5).$  Since the treedepth of $\overline{T}$ is bounded, $a$ and $b$ have only a few ancestors. Thus, only a few table lookups will be done in the call to $\texttt{ExplodedCallGraphReachability}.$ Finally, when we reach the beginning of our target function $f_j,$ we have to take an SCVP to our target state $(u_2, d_2).$ To check if such a path exists, we simply rely on the same-context queries of Chapter~\ref{sec:review}.

\paragraph{Runtime Analysis of the Preprocessing Phase} Our algorithm is much faster than the classical IFDS algorithm of~\cite{reps}. More specifically, for the preprocessing, we have:
\begin{itemize}
	\item Step 1 is the same as Chapter~\ref{sec:review} and takes $O(n \cdot D^3)$ time.
	\item Step 2 is a simple intraprocedural analysis that runs a reverse DFS from every node $(c, d)$ in any function $f$. Assuming that the function $f$ has $\alpha$ lines of code and a total of $\delta$ function call statements, this will take $O(\alpha \cdot \delta \cdot D^3).$ Assuming that $\delta$ is a small constant, this leads to an overall runtime of $O(n \cdot D^3).$ This is a realistic assumption since we rarely, if ever, encounter functions that call more than a constant number of other functions.
	\item In Step 3, we have at most $O(n \cdot D)$ call nodes of the form $(c, d_3).$ Based on the bounded bandwidth assumption, each such node leads to constantly many possibilities for $d_2$. So, we perform at most $O(n \cdot D^2)$ calls to the same-context query procedure. Each same-context query takes $O(\lceil D / \lg n \rceil),$ so the total runtime of this step is $O(n \cdot D^3 / \lg n).$
	\item In Step 4, the total time for computing all the $up$ and $down$ values is $O(n \cdot D^3 \cdot k_2).$ This is because $\overline{C}$ has at most $O(n \cdot D)$ vertices and $O(n \cdot D^2)$ edges and each edge can be traversed at most $O(D \cdot k_2)$ times in the DFS, where $k_2$ is the depth of our POT for $C$. The treedepth of $\overline{C}$ is a factor $D$ larger than that of $C.$
\end{itemize}
Putting all these points together, the total runtime of our preprocessing phase is $O(n \cdot D^3 \cdot k_2),$ which has only linear dependence on the number of lines, $n$.

\paragraph{Runtime Analysis of the Query Phase} To analyze the runtime of a query, note that there are $O(\delta \cdot D)$ different possibilities for $(c, d_3).$ Due to the bounded bandwidth assumption, each of these corresponds to a constant number of $(f_x, d_4)$'s. For each $(f_x, d_4)$ and $(f_j, d_5),$ we perform a reachability query using the POT $\overline{T},$ in which we might have to try up to $O(k_2 \cdot D)$ common ancestors. So, the total runtime for finding all the reachable $(f_2, d_5)$'s from all $(f_x, d_4)$'s is $O(D^3 \cdot k_2 \cdot \delta).$ Finally, we have to perform a same-context query from every $(s_j, d_5)$ to $(u_2, d_2).$ So, we do a total of at most $O(D)$ queries, each of which takes $O(\lceil D / \lg n \rceil ),$ and hence the total runtime is $O(D^3 \cdot k_2 \cdot \delta),$ which is $O(D^3)$ in virtually all real-world scenarios where $k_2$ and $\delta$ are small constants.

\chapter{Experimental Results} \label{sec:exper}

\paragraph{Implementation and Machine} We implemented the algorithms of Chapters~\ref{sec:review} and~\ref{sec:algo}, as well as the approaches of~\cite{reps} and~\cite{demand}, in a combination of C++ and Java, and used the Soot framework~\cite{soot} to obtain the control-flow and call graphs. Specifically, we use the SPARK call graph created by Soot for the intermediate Jimple representation. To compute treewidth and treedepth, we used the winning open-source tools submitted to past PACE challenges~\cite{dell_et_al:LIPIcs:2018:8558,kowalik_et_al:LIPIcs:2020:13340}. All experiments were run on an Intel i7-11800H machine (2.30 GHz, 8 cores, 16 threads) with 12 GB of RAM.

\paragraph{Benchmarks and Experimental Setup} We compare the performance of our method against the standard IFDS algorithm~\cite{reps} and its on-demand variant~\cite{demand} and use the standard DaCapo benchmarks~\cite{dacapo} as input programs. These are real-world programs with hundreds of thousands of lines of code. For each benchmark, we consider three different classical data-flow analyses: (i) reachability analysis for dead-code elimination, (ii) null-pointer analysis, and (iii) possibly-uninitialized variables analysis. For each analysis, we gave each of the algorithms 10 minutes time over each benchmark and recorded the number of queries that the algorithm successfully handled in this time. The queries themselves were randomly generated\footnote{For generating each query, we randomly and uniformly picked two points in the exploded supergraph. Note that none of our queries are same-context. Even when the two points of the query are in the same function, we are asking for reachability using interprocedurally valid paths that are not necessarily same-context.} and the number of queries was also limited to $n$, i.e.~the number of lines in the code. We then report the average cost of each query, i.e.~each algorithm's total runtime divided by the number of queries it could handle. The reason for this particular setup is that~\cite{reps} and~\cite{demand} do not distinguish between preprocessing and query. So, to avoid giving our own method any undue advantage, we have to include both our preprocessing and query time in the mix.

\paragraph{Treewidth and Treedepth} In our experiments, the maximum encountered treewidth was $10,$ whereas the average was $9.1$. Moreover, the maximum treedepth was $135$ and the average was $43.8$. Hence, our central hypothesis that real-world programs have small treewidth and treedepth holds in practice and the widths and depths are much smaller than the number of lines in the program. See table~\ref{tab:twtd}.

\begin{table}[!ht]
    \centering
    \begin{tabular}{|l|l|l|}
    \hline
        Benchmark & Treewidth & Treedepth  \\ \hline
        hsqldb & 7 & 6  \\ \hline
        xalan & 7 & 6  \\ \hline
        avrora & 9 & 15  \\ \hline
        fop & 8 & 13  \\ \hline
        luindex & 9 & 17  \\ \hline
        lusearch & 10 & 16  \\ \hline
        eclipse & 10 & 29  \\ \hline
        antlr & 10 & 46  \\ \hline
        pmd & 9 & 53  \\ \hline
        sunflow & 10 & 102  \\ \hline
        jython & 10 & 67  \\ \hline
        chart & 9 & 65  \\ \hline
        bloat & 10 & 135  \\ \hline
    \end{tabular}
    \caption{The treewidth and treedepth for each benchmark in our experiments. Treewidth denotes the maximum treewidth among all control-flow graphs in a benchmark.}
    \label{tab:twtd}
\end{table}

\begin{figure}
	\begin{center}
		\includegraphics[keepaspectratio,scale=0.35]{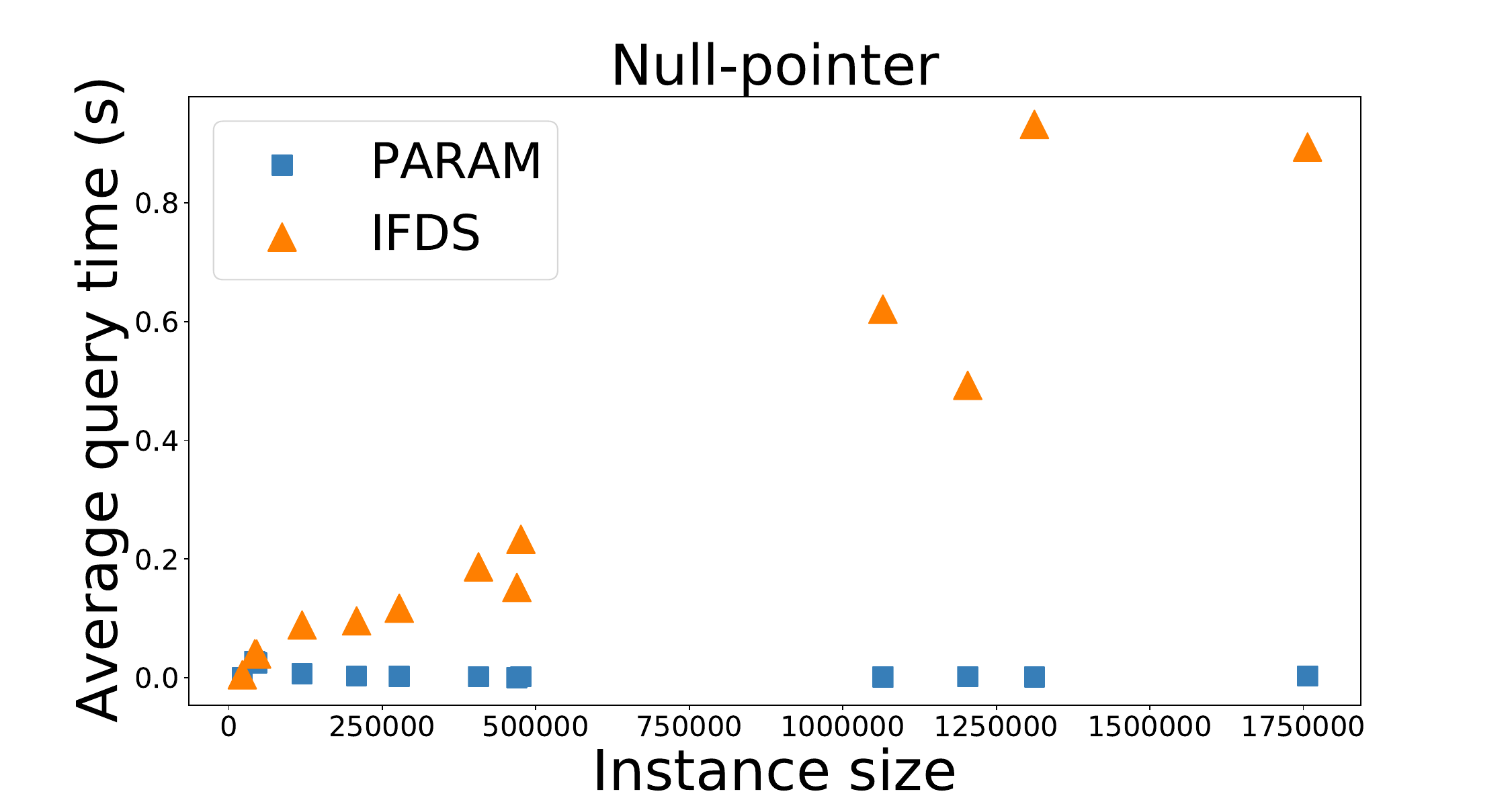}\\
		\includegraphics[keepaspectratio,scale=0.35]{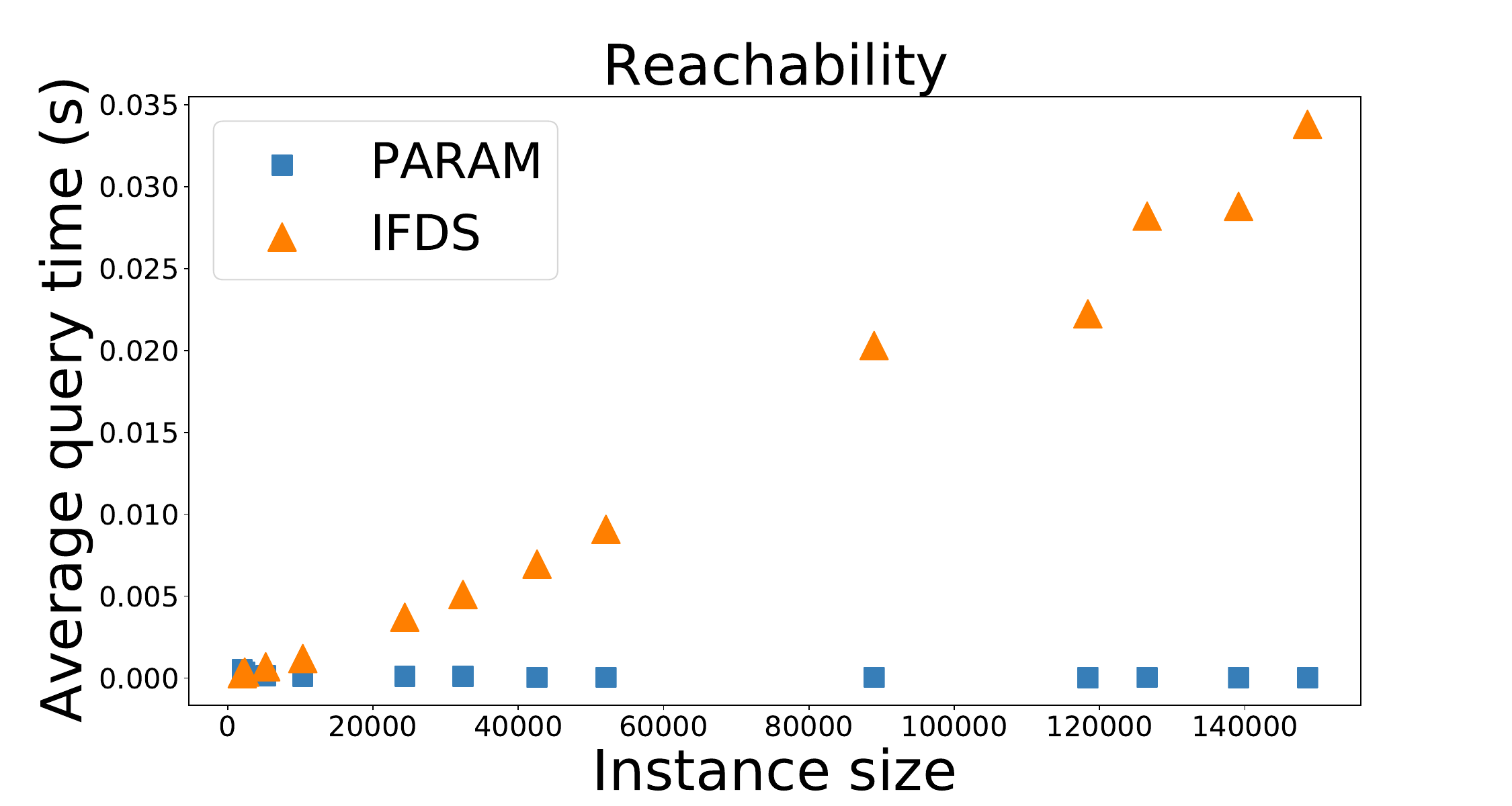}\\
		\includegraphics[keepaspectratio,scale=0.35]{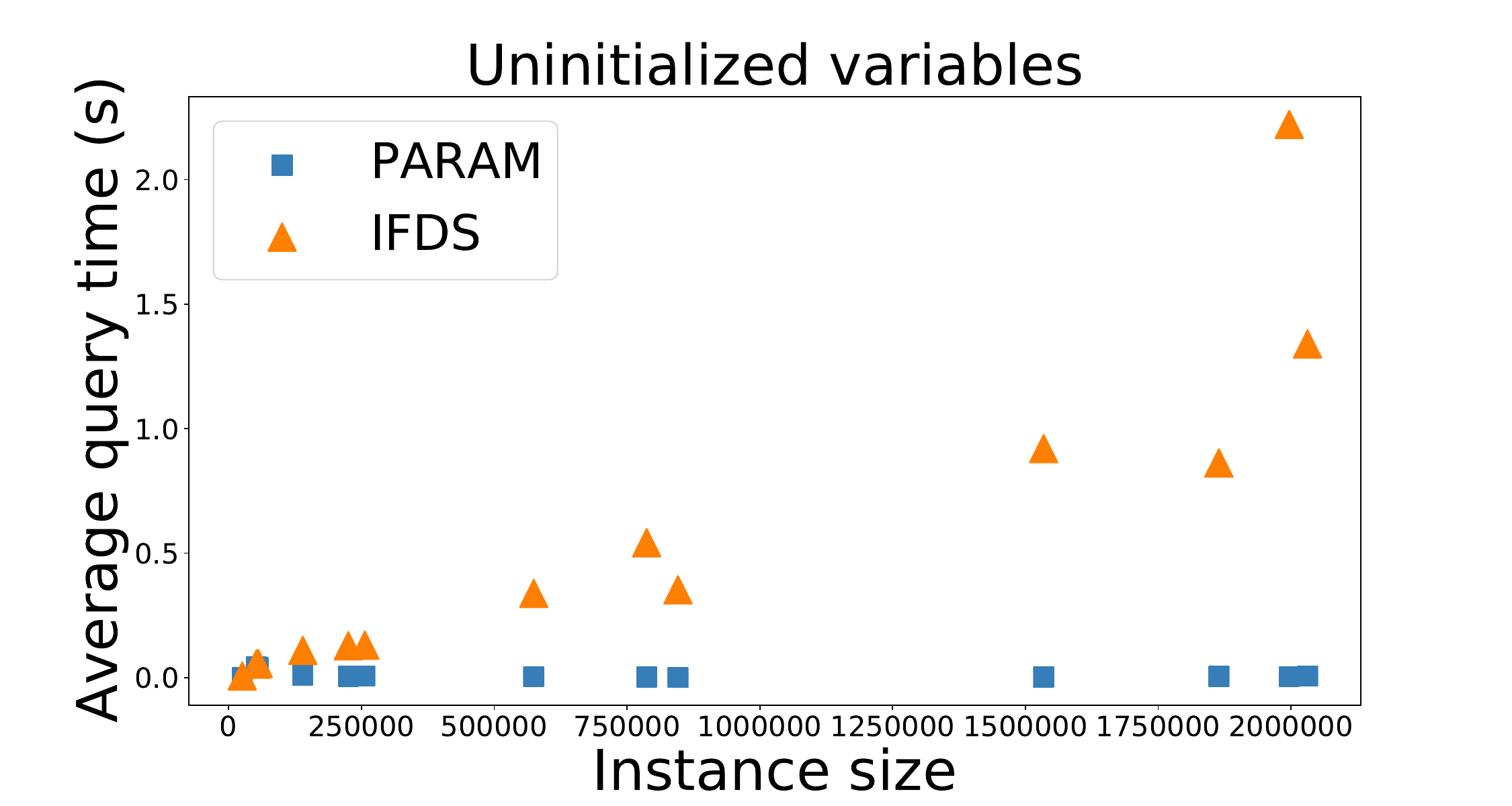}
	\end{center}
\caption{Comparison of the average cost per query for our algorithm vs~\cite{reps}.}
\label{fig:exper-ifds}
\end{figure}

\begin{figure}
	\begin{center}
		\includegraphics[keepaspectratio,scale=0.35]{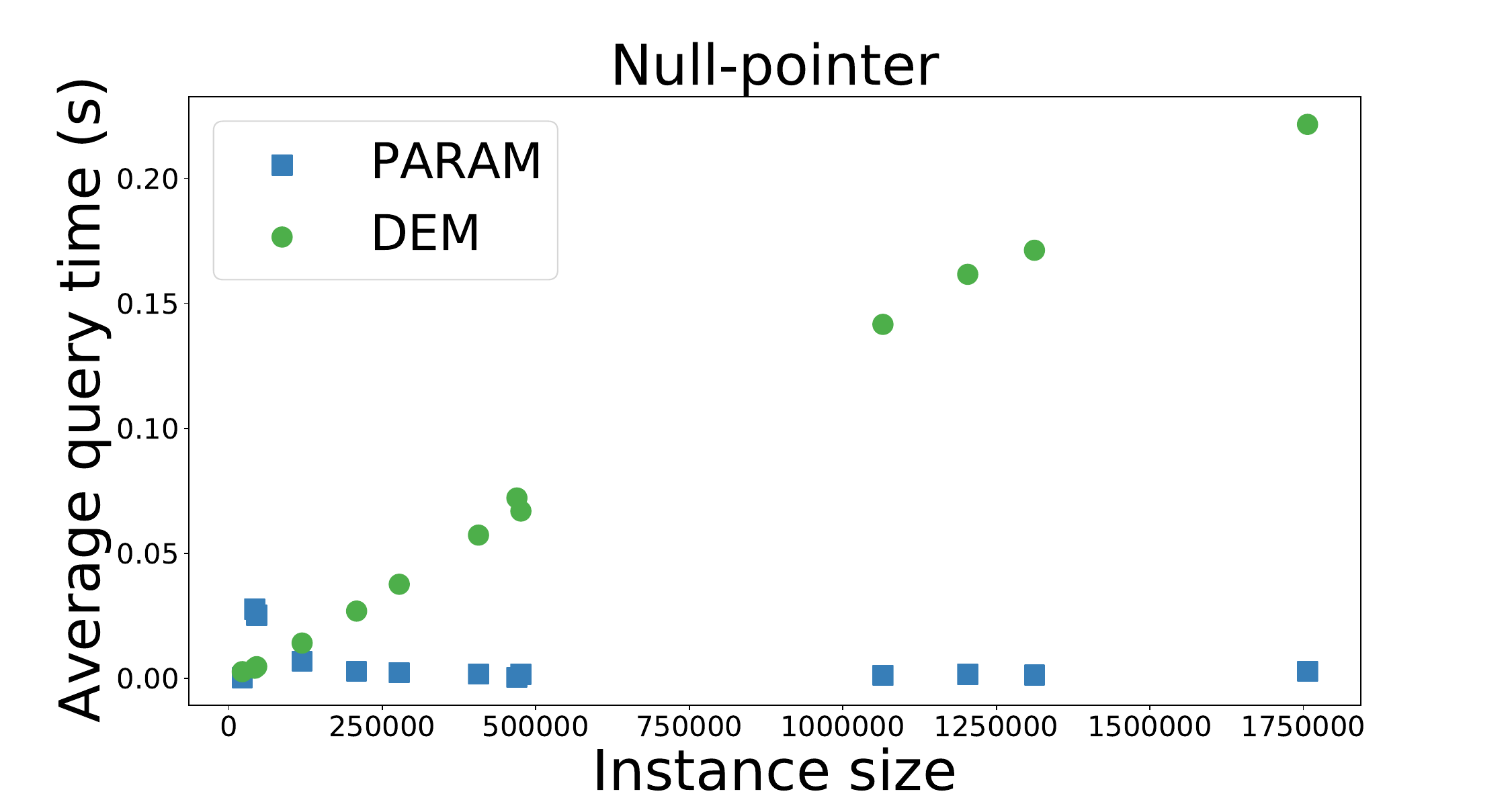}\\
		\includegraphics[keepaspectratio,scale=0.35]{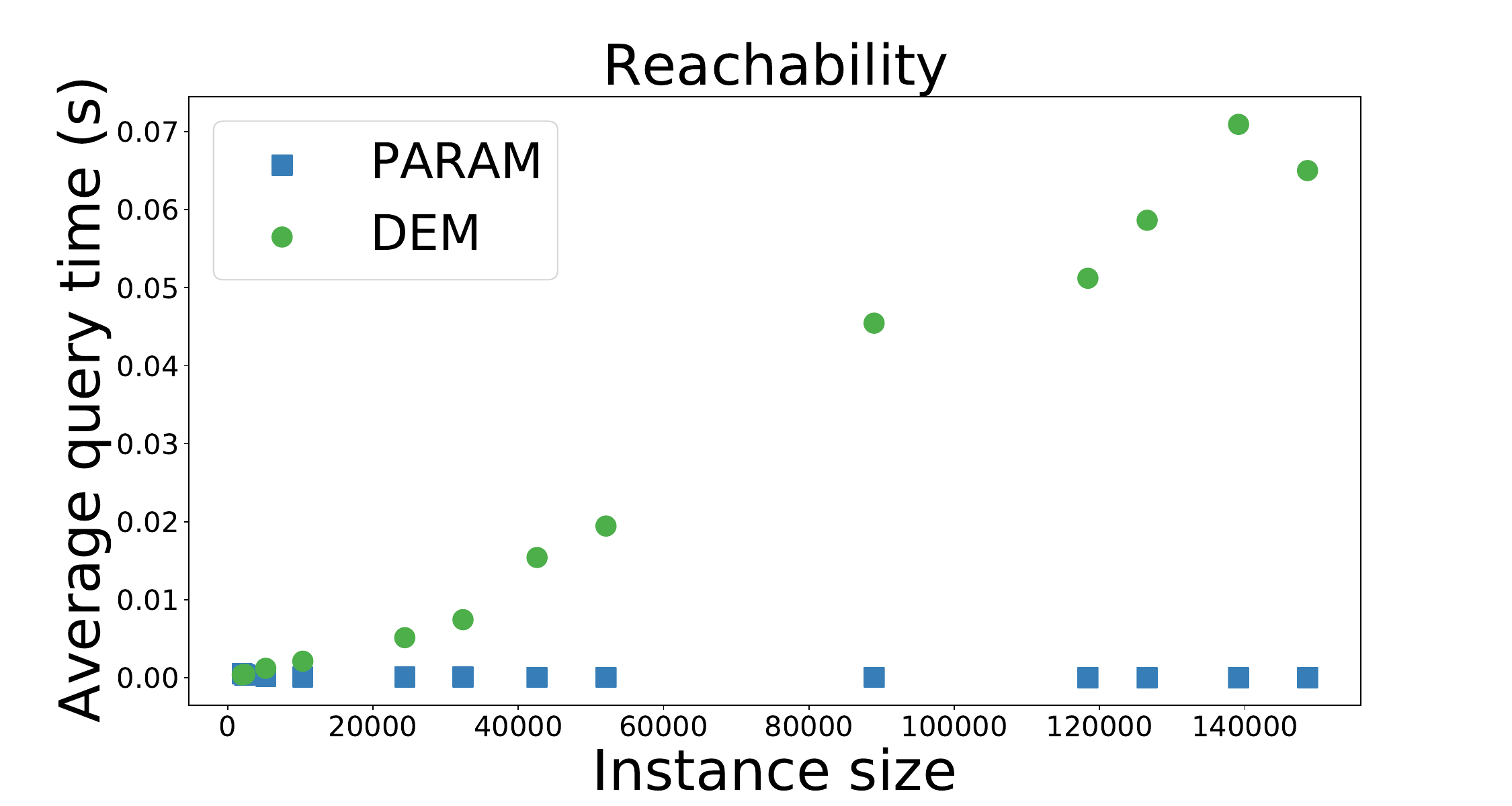}\\
		\includegraphics[keepaspectratio,scale=0.35]{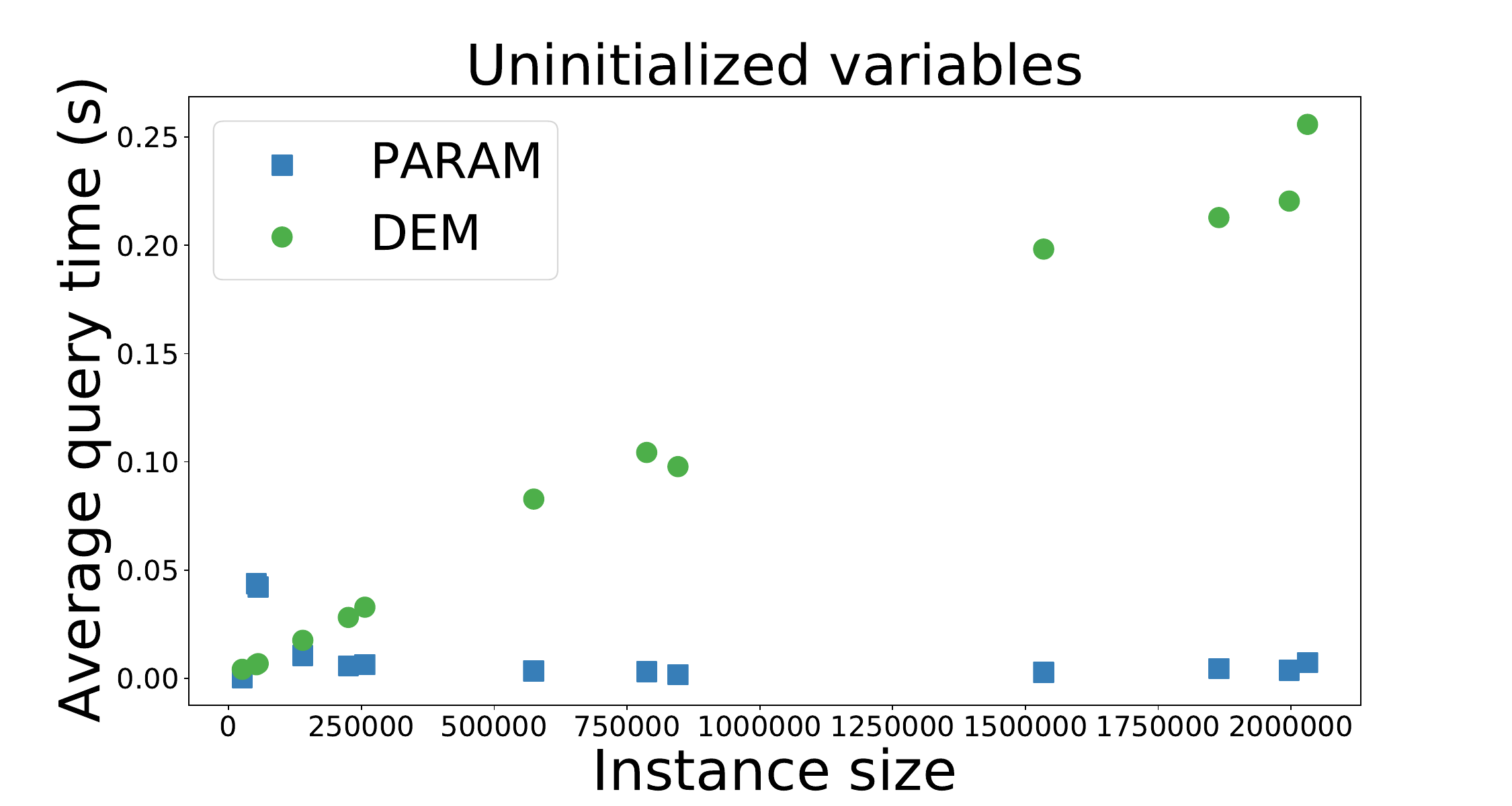}
	\end{center}
\caption{Comparison of the average cost per query for our algorithm vs~\cite{demand}.}
\label{fig:exper-dem}
\end{figure}

\paragraph{Results} Table~\ref{tab:big} shows our experiments results with one row for each analysis and benchmark. The graphs in Figure~\ref{fig:exper-ifds} and Figure~\ref{fig:exper-dem} provide the average query time for each analysis. Each dot corresponds to one benchmark. We use \texttt{PARAM}, \texttt{IFDS} and \texttt{DEM} to refer to our algorithm, the IFDS algorithm in~\cite{reps}, and the on-demand IFDS algorithm in~\cite{demand}, respectively. The reported instance sizes are the number of edges in $\overline{G}$.

\begin{table}
\centering
\label{tab:big}
\begin{tabular}{|l|l|l|l|l|l|l|l|l|} 
\hline
Analysis                                                                                                  & BM       & $|\overline{G}|$ & $\texttt{Prec}$ & $\texttt{PARAM}$ & $\texttt{IFDS}$ & $\texttt{DEM}$ & $\texttt{I/P}$ & $\texttt{D/P}$  \\ 
\hline
\multirow{13}{*}{\begin{tabular}[c]{@{}l@{}}Reachability\\analysis\end{tabular}}                          & hsqldb   & 2015             & 0.48            & 0.54             & 0.27            & 0.37           & 0.5            & 0.7             \\
                                                                                                          & xalan    & 2357             & 0.38            & 0.36             & 0.36            & 0.44           & 1.0            & 1.2             \\
                                                                                                          & avrora   & 5244             & 0.29            & 0.14             & 0.69            & 1.21           & 5.0            & 8.9             \\
                                                                                                          & fop      & 10352            & 0.38            & 0.09             & 1.19            & 2.12           & 14.0           & 25.0            \\
                                                                                                          & luindex  & 24382            & 1.11            & 0.11             & 3.71            & 5.14           & 34.0           & 47.1            \\
                                                                                                          & lusearch & 32393            & 1.31            & 0.10             & 5.10            & 7.45           & 52.1           & 76.1            \\
                                                                                                          & eclipse  & 42583            & 0.68            & 0.04             & 6.95            & 15.42          & 177.1          & 392.7           \\
                                                                                                          & antlr    & 52069            & 0.68            & 0.04             & 9.08            & 19.45          & 250.2          & 536.1           \\
                                                                                                          & pmd      & 88968            & 1.13            & 0.03             & 20.30           & 45.45          & 589.4          & 1319.5          \\
                                                                                                          & sunflow  & 118389           & 1.16            & 0.03             & 22.24           & 51.20          & 879.6          & 2024.9          \\
                                                                                                          & jython   & 126544           & 1.52            & 0.03             & 28.20           & 58.63          & 885.4          & 1840.6          \\
                                                                                                          & chart    & 139125           & 1.50            & 0.03             & 28.80           & 70.92          & 1008.7         & 2483.9          \\
                                                                                                          & bloat    & 148616           & 1.54            & 0.03             & 33.81           & 65.00          & 1180.2         & 2269.1          \\ 
\hhline{|=========|}
\multirow{13}{*}{\begin{tabular}[c]{@{}l@{}}Null-pointer\\analysis\end{tabular}}                          & avrora   & 22040            & 0.69            & 0.32             & 5.01            & 2.69           & 15.5           & 8.4             \\
                                                                                                          & hsqldb   & 42451            & 24.75           & 27.64            & 40.42           & 4.08           & 1.5            & 0.1             \\
                                                                                                          & xalan    & 45536            & 26.63           & 25.27            & 40.05           & 4.69           & 1.6            & 0.2             \\
                                                                                                          & fop      & 119515           & 26.89           & 6.83             & 88.81           & 14.14          & 13.0           & 2.1             \\
                                                                                                          & luindex  & 208266           & 28.68           & 2.85             & 95.87           & 26.92          & 33.7           & 9.5             \\
                                                                                                          & lusearch & 277691           & 31.49           & 2.35             & 117.16          & 37.65          & 49.9           & 16.0            \\
                                                                                                          & eclipse  & 406720           & 29.82           & 1.74             & 186.83          & 57.31          & 107.1          & 32.9            \\
                                                                                                          & pmd      & 469261           & 11.88           & 0.35             & 152.17          & 72.16          & 432.9          & 205.3           \\
                                                                                                          & antlr    & 475751           & 31.75           & 1.66             & 233.25          & 66.93          & 140.4          & 40.3            \\
                                                                                                          & jython   & 1065292          & 53.66           & 1.27             & 620.94          & 141.59         & 488.8          & 111.5           \\
                                                                                                          & sunflow  & 1203363          & 72.74           & 1.63             & 492.55          & 161.61         & 303.1          & 99.4            \\
                                                                                                          & chart    & 1312019          & 63.59           & 1.28             & 932.04          & 171.24         & 729.3          & 134.0           \\
                                                                                                          & bloat    & 1756723          & 140.86          & 2.78             & 893.82          & 221.58         & 321.2          & 79.6            \\ 
\hhline{|=========|}
\multirow{13}{*}{\begin{tabular}[c]{@{}l@{}}Possibly-\\uninitialized \\variables \\analysis\end{tabular}} & avrora   & 25794            & 0.73            & 0.34             & 6.88            & 4.15           & 20.2           & 12.2            \\
                                                                                                          & hsqldb   & 52342            & 39.14           & 43.70            & 54.40           & 6.28           & 1.2            & 0.1             \\
                                                                                                          & xalan    & 55747            & 44.36           & 42.12            & 57.18           & 6.78           & 1.4            & 0.2             \\
                                                                                                          & fop      & 139747           & 41.63           & 10.56            & 109.79          & 17.55          & 10.4           & 1.7             \\
                                                                                                          & luindex  & 225449           & 39.61           & 5.74             & 128.29          & 28.13          & 22.3           & 4.9             \\
                                                                                                          & lusearch & 256597           & 52.10           & 6.29             & 130.64          & 32.86          & 20.8           & 5.2             \\
                                                                                                          & eclipse  & 574471           & 55.77           & 3.34             & 338.28          & 82.79          & 101.3          & 24.8            \\
                                                                                                          & antlr    & 787078           & 59.63           & 3.11             & 542.31          & 104.32         & 174.2          & 33.5            \\
                                                                                                          & pmd      & 845938           & 51.99           & 1.74             & 353.17          & 97.78          & 202.5          & 56.1            \\
                                                                                                          & jython   & 1534301          & 114.05          & 2.68             & 920.00          & 198.25         & 342.8          & 73.9            \\
                                                                                                          & bloat    & 1864126          & 214.35          & 4.53             & 862.59          & 212.82         & 190.3          & 47.0            \\
                                                                                                          & chart    & 1996635          & 184.09          & 3.72             & 2221.90         & 220.44         & 597.8          & 59.3            \\
                                                                                                          & sunflow  & 2030993          & 265.32          & 7.20             & 1340.47         & 255.87         & 186.2          & 35.6            \\
\hline
\end{tabular}
\caption{Our experimental results. BM denotes the benchmark, $|\overline{G}|$ denotes the number of edges in $\overline{G},$ \texttt{Prec} is our preprocessing time in seconds, \texttt{PARAM}, \texttt{IFDS}, and \texttt{DEM} denote the average query runtime for the corresponding algorithm in milliseconds. \texttt{I/P} is the ratio of \texttt{IFDS} to \texttt{PARAM} and similarly \texttt{D/P} is the ratio of \texttt{DEM} to \texttt{PARAM}.}
\end{table}


\paragraph{Discussion} As shown in Chapter~\ref{sec:algo}, our algorithm's preprocessing has only linear dependence on the number $n$ of lines and our query time is completely independent of $n$. Thus, our algorithm has successfully pushed most of the time complexity on the small parameters such as the treewidth $k_1$, treedepth $k_2$, bandwidth $b$ and maximum number of function calls in each function, i.e.~$\delta.$ All these parameters are small constants in practice. Specifically, the two most important ones are always small: The treewidth in DaCapo benchmarks never exceeds $10$ and the treedepth is at most $135$. This is in contrast to $n$ which is the hundreds of thousands and the instance size, which can be up to around $2 \cdot 10^6.$ In contrast, both~\cite{reps} and~\cite{demand} have a quadratic dependence on $n$. Unsurprisingly, this leads to a huge gap in the practical runtimes and our algorithm is on average faster than the best among~\cite{reps} and~\cite{demand} by a factor of $158,$ i.e.~more than two orders of magnitude. Moreover, the difference is much starker on larger benchmarks, in which the ratio of our parameters to $n$ is close to $0$. On the other hand, in a few small instances, simply computing the treewidth and treedepth is more time-consuming than the previous approaches and thus they outperform us.
\chapter{Conclusion} \label{sec:conclusion}

In this work, we provided a parameterized algorithm for the general case of on-demand inteprocedural data-flow analysis as formalized by the IFDS framework. We exploited a novel parameter, i.e.~the treedepth of call graphs, to reduce the runtime dependence on the number of lines of code from quadratic to linear. This led to significant practical improvements of more than two orders of magnitude in the runtime of the IFDS data-flow analysis as demonstrated by our experimental results. Moreover, this is the first theoretical improvement in the runtime of the general case of IFDS since the original algorithm of~\cite{reps}, which was published in 1995. In contrast, previous approaches, such as~\cite{esop}, could only improve the runtime for same-context queries.

\newpage
\addcontentsline{toc}{chapter}{References}
\bibliographystyle{IEEEtran}
\bibliography{refs}

\end{document}